\documentclass[titlepage,12pt]{utarticle}

\usepackage{amsmath}
\usepackage{mathrsfs}
\long\def\omit#1{}

\newcommand{\dd}{\mathrm{d}}
\newcommand\HH{\mathcal{H}}

\newcommand\MM{\mathcal{M}}
\newcommand\RR{\mathcal{R}}
\newcommand\OO{\mathcal{O}}
\newcommand\DD{\mathcal{D}}
\newcommand\GG{\mathcal{G}}

\newcommand\dM{\partial \MM}

\newcommand\FF{\mathcal{F}}

\newcommand\eps{\epsilon}

\newcommand\nts{\negthickspace}
\newcommand\bns{\nts \nts \nts}

\newcommand\tg{\tilde{g}}

\newcommand\vf{\varphi}
\newcommand\dvf{\delta \vf}
\newcommand\pvf{\partial_{\vf}}

\DeclareMathAlphabet{\mathpzc}{OT1}{pzc}{m}{it}
\newcommand\Ham{\mathscr{H}}
\newcommand\Lag{\mathscr{L}}

\begin{document}
%\sffamily

\preprint{MCTP--03-32\\ {\tt hep-th/0307026}\\ }

\title{Inflation and de~Sitter Holography}

\author{Finn Larsen\footnote{\texttt{larsenf@umich.edu}} ~and Robert McNees\footnote{\texttt{ramcnees@umich.edu}}}

\oneaddress{Michigan Center for Theoretical Physics\\
  University of Michigan\\
        Ann Arbor, MI-48109, USA  }

\date{}

\Abstract{ 
We develop the relation between de~Sitter holography and inflation in detail
with particular attention to cosmic density perturbations. 
We set up the Hamilton-Jacobi formalism to present a systematic treatment of the 
logarithmic corrections to a scale invariant spectrum. Our computations can be 
interpreted without reference to holography, as strong infra-red effects in gravity. 
This point of view may be relevant for the fine-tuning problems inherent to inflation.
}

\maketitle

\section{Introduction} 
\label{sec:introduction}

Inflation is the leading candidate for explaining how causal physics in the early universe 
produced the large scale structure we observe today. One of its most successful predictions 
is the existence of an approximately scale-invariant spectrum of anisotropies
in the cosmic microwave background radiation \cite{Bardeen:qw,Mukhanov:xt,Mukhanov:nu,Starobinsky:ee,
Guth:ec,Mukhanov:1990me}. These anisotropies are now being observed 
in impressive detail \cite{Bennett:2003bz}, yielding a precise picture of the slight deviations 
from scale-invariance in the CMBR. 

While inflation is a phenomenologically succesful paradigm, it is highly unsatisfying
theoretically. The inflationary potential must be chosen to have a minimum extremely 
close to zero, or else superluminal expansion would not end. Additionally, the potential driving 
the expansion must be chosen exceptionally flat, in order to inflate for sufficiently long. These 
requirements amount to the introduction of exceedingly small 
dimensionless parameters in the effective Lagrangian describing the inflationary epoch 
\cite{Lyth:1998xn,Riotto:2002yw}. This undermines the separation of scales that we
expect in effective field theory. Thus inflation suggests that, when gravity is taken into account, the standard interpretation 
of effective quantum field theory needs some modification. The puzzles posed by dark 
energy similarly indicate that quantum gravity holds secrets already at low energy.

The purpose of this paper is to present some computations which seem relevant to these issues.
The technical question we solve is to isolate the infrared divergences of classical fields in 
inflationary spacetimes. We consider spectator scalars, as well as the full gravity-scalar
system. In both cases we compute a renormalized effective action, with infrared divergences 
removed. The remaining anomalous scaling behavior
governs departures from scale invariance in the power spectrum of the primordial density perturbations. 
We find results that agree with the standard inflationary predictions, but our computation is organized very differently.

As we know from standard quantum field theory, without gravity, scaling solutions are fixed 
points in the space of theories. As such, they exhibit many universal features. Our framework 
could therefore be helpful in identifying the correct theory of inflation. More importantly, 
fine-tuning issues should clearly be understood in terms of scaling behavior, so our 
considerations seem relevant for these notorious problems. We are certainly not yet able
to address the fine-tuning problems. Our point is simply to stress the central lesson from quantum 
field theory that divergences matter, technically and conceptually. This makes a systematic 
treatment of infrared divergences in inflation worthwhile, perhaps even essential.

In our computations we consider the spacetime action at some late time $\tau$ as a functional 
of the scalar field $\vf(\vec{x})$ that time. A practical way to do that is to employ the Hamilton-Jacobi
formalism and identify the spacetime action
%, and so the semi-classical wave-function of the universe, 
with the Hamilton-Jacobi functional $S(\vf,\tau)$. It is this quantity, which we can also interpret as the phase
of the semi-classical wave function of the Universe, which suffers
infra-red divergences. After removing all local divergences we are left with certain logarithmic
terms which control the scaling properties of the CMBR. Applications of the Hamilton-Jacobi 
formalism to inflation have been studied extensively in the literature \cite{Salopek:1990jq, Salopek:1994sq,
Salopek:1997he, Parry:mw}.

Our approach is heavily motivated by considerations of holography. Some interpretations of holography
suggest that a gravitational theory in four dimensions can be `dual' to a local, 
non-gravitational theory in three
dimensions. This approach to holography implies a correspondence between infrared physics in one theory and
ultraviolet physics in the dual theory. In this sense, the infrared divergences we study in the gravitational
theory would correspond to ultraviolet divergences in the dual three dimensional theory. Violations of scale
invariance in the four dimensional theory are then tied to the usual structures one employs to study departures
from scale invariance in a local, non-gravitational theory, such as $\beta$-functions and anomalous dimensions \cite{Larsen:2002et}.
Although we are very sympathetic to the ideas surrounding a holographic interpretation, we will not emphasize holography
throughout most of this paper. We will instead adopt a more conservative point of view: that we are simply discussing
infrared divergences of gravity, in the belief that they play a central role in theories of inflation. Of course, we
cannot fully resist discussing some aspects of the holographic dictionary alluded to above; we do so in section 6.

It is worth noting that our approach differs technically from that of the conjectured dS/CFT correspondence 
\cite{Strominger:2001pn,Strominger:2001gp, Spradlin:2001nb,Tolley:2001gg}. We do not consider global de Sitter space, rather we restrict
our attention to the ``late time" Poincare patch. The boundary conditions at early times are determined by our
insistence that the analytically continued theory is regular. This is reminiscent of AdS/CFT (as opposed to dS/CFT) 
and amounts to having no incoming radiation at early times, as is customary and physically appropriate in
cosmology. In this sense our calculations are similar to the variation of dS/CFT proposed by Maldacena~\cite{Maldacena:2002vr}.

We should also comment on the relation to the holographic renormalization group~\cite{deBoer:1999xf,Bianchi:2001kw,
Bianchi:2001de,DeWolfe:2000xi, Freedman:1998tz, Balasubramanian:1999jd}. This term usually refers to scenarios
where our 4D universe is embedded in a 
higher dimensional curved space time, with one of the extra dimensions admitting an alternate interpretation as a
renormalization group scale in the 4D theory. In contrast, to the extent we interpret our 
results holographically, we consider cosmological evolution a flow in the space of 3D
theories.
Despite this difference in perspective, we find much technical overlap with several works on the 
holographic renormalization group, particularly~\cite{deBoer:1999xf}. 
One notable difference is that we consider bulk gravity in an even dimensional spacetime. This
is in contrast to the standard examples in the holography literature, involving odd dimensional
spacetimes such as $(A)dS_5$ or $(A)dS_3$.

The paper is organized as follows.
In section \ref{sec:HJTheory} we briefly review Hamilton-Jacobi theory, particularly
the introduction of the classical action, the Hamilton-Jacobi functional. 
In section \ref{sec:strategy} we compute the Hamilton-Jacobi functional  
for a spectator scalar and discuss the interpretation of its infra-red divergences.
Section \ref{sec:fluctuations} contains the explicit computation of the spectrum
of fluctuations of a massive scalar field in a fixed de~Sitter background. This allow 
us to give an example of the
formalism in a setting which is relatively simple because it neglects the 
technical complications due to gravitational backreaction. The more complete
case, including gravitational backreaction, is considered in section \ref{sec:gravscalar}.
In section \ref{sec:rgcft} we discuss the implications for the interpretation of inflation
as broken scale invariance, in the spirit of holography.
The appendices contain conventions as well as the details of some of the calculations 
contained in section \ref{sec:gravscalar}.

\section{Hamilton-Jacobi Theory} \label{sec:HJTheory}
The Hamilton-Jacobi formalism provides a powerful framework for solving
problems in classical mechanics. It also provides a natural intermediate step between 
classical and quantum mechanics, because it governs the phase of the wave function 
in the semi-classical approximation. 
The H-J formalism will be of central importance in our considerations so we begin
with a short review. 

\subsection{Mechanics}
Consider a classical system with a single dynamical variable $q$. The action is
written in terms of the Lagrangian as: 
\begin{equation}
	S =\int_{t_i}^{t_f} \nts \nts dt~ L(q,\dot{q},t) 
	\label{Scldef}
\end{equation}
%The Hamilton-Jacobi function is defined as follows. It is the action $S$ but it is interpreted
%in an unfamiliar way, as a function of the time $t_f$ and the value of the dynamical variable 
%at that time, {\it i.e.} $q(t_f)$. 
The Hamilton-Jacobi function is defined as the classical action $S$, interpreted as a function of
the time $t_f$ and the value of the dynamical variable at that time, $q(t_f)$. Here ``classical" means
that the action should be evaluated with the initial value $q(t_i)$ fixed and $q(t_f)$ ``on-shell", {\it i.e.} satisfying
its equation of motion. We write the H-J function as $S=S(q,t)$ with the understanding that $t=t_f$ 
and $q=q(t_f)$.

The variation of the H-J function (\ref{Scldef}) with respect to $q$ is:
\begin{equation}
\delta_q S(q,t) =\left.  {\partial L\over \partial \dot{q}}\delta q \right|^{t_f}_{t_i}
+\int^{t_f}_{t_i}
 dt \left( {\partial L\over \partial q}- {d\over dt}{\partial L\over \partial \dot{q}}\right) \delta q
\label{svar}
\end{equation}
after integration by parts. The integrand is proportional to the equation of motion, which vanishes
when $q(t)$ is on-shell. Since the initial value of $q(t_i)$ is fixed $\delta q(t_i) = 0$, and $\delta S$
depends only on $\delta q(t_f)$. Thus, as anticipated, $S$ is an ordinary 
function of $q=q(t_f)$, rather than a functional of $q(t)$. Its derivative is simply:
\begin{equation}
{\partial S\over \partial q}={\partial L\over \partial \dot{q}} = p
\label{pcan}
\end{equation}
where $p$ is the canonical momentum.

The H-J function depends  on time explicitly, as well as through $q(t)$. We can infer
this dependence by taking the total derivative of  (\ref{Scldef}) and expanding as:
\begin{equation}
{dS\over dt} = L = {\partial S\over \partial t} + {\partial S\over \partial q}\dot{q}
\end{equation}
Reorganizing this equation using \eqref{pcan} we find the Hamiltonian:
\begin{equation}
H =  p\dot{q} - L = -{\partial S\over \partial t} 
\label{hamdef}
\end{equation}
The Hamiltonian is defined as a function of coordinates and momenta $H=H(q,p)$ but
we can use the expression \eqref{pcan} for the momentum and so:
\begin{equation}
{\partial S\over \partial t} + H(q,{\partial S\over \partial q})=0
\label{hjeqn}
\end{equation}
For a given dynamical system the Hamiltonian is a specific function and \eqref{hjeqn}
becomes a powerful first order differential equation for the H-J function $S(q,t)$. 
It is known as the Hamilton-Jacobi equation.

The connection with the semiclassical approximation to quantum mechanics appears when we consider the 
Schr\"{o}dinger equation for the wave function $\psi(q)$:
\begin{eqnarray} \label{schrodinger}
	i \hbar \frac{\partial}{\partial \, t} \psi(q) & = & H\left(q, -i\hbar \frac{\partial}{\partial q}  \right) 
		\psi(q)
\end{eqnarray}
Taking the saddle-point approximation:
\begin{equation}\label{scwf}
 	\psi(q) \sim \exp\left(\frac{i}{\hbar} S(q,t)\right) 
\end{equation}
the Schr\"{o}dinger equation \eqref{schrodinger} reduces to the Hamilton-Jacobi equation \eqref{hjeqn}.
Thus the H-J function can be interpreted as the phase of the wave-function in the
semi-classical approximation.

\subsection{Field Theory}
The generalization of these considerations to fields is straightforward. The H-J function
is now a function of the time $\tau$ and a \emph{functional} of the field configuration $\vf(\vec{x},\tau)$ 
evaluated at that time. We will often refer to $\vf(\vec{x},\tau)$ as the `boundary data' for the field $\vf$.

Our main interest is when fields are coupled to gravity. To keep things simple we defer the
discussion of the general gravity-scalar system to section \ref{sec:gravscalar}.
For now we consider the case of a scalar field on a fixed background spacetime. The metric takes the 
cosmological form:
\begin{equation}
	\dd s^2 =  a(\tau)^2 \, \left( - \dd \tau^2 + \gamma_{ij}(\vec{x})\dd x^i \dd x^j\right)
\end{equation}
Note that we use the conformal time $\tau$, related to the more
conventional time $t$ through $\dd t = a(\tau)\dd \tau$. In this background the 
Lagrangian for a scalar field takes the form:
\begin{equation}
\label{simplecalaraction}
\Lag  =  
  \frac{1}{2} \left( \frac{\vf'}
			{a}\right)^2 - \frac{1}{2} \, \vec{D} \vf \cdot  \vec{D} \vf - V(\vf) 
\end{equation}
where the prime denotes derivative with respect to the conformal time $\tau$ 
and spatial indices are contracted using the full spatial
metric $\tilde{g}_{ij}=a(\tau)^2\gamma_{ij}$
(or more precisely its inverse). Let us also record the corresponding Hamiltonian density: 
\begin{equation}
\label{Hdensity}
	\Ham  =  \frac{1}{2} \, \pi^2 + \frac{1}{2} \vec{D} \vf \cdot \vec{D} \vf + V(\vf)
\end{equation}
where the momentum is:
\begin{equation}
\label{simplemom}
\pi =  \vf^\prime / a 
\end{equation}

Now, as in simple mechanics, the linchpin of the Hamilton-Jacobi formalism
is the action:
\begin{equation}\label{firstscalaraction}
	S  =  \int_{\tau_i}^{\tau_f} \nts \nts\dd^3 x \, \dd\tau  \sqrt{-g} \, \Lag \left( \vf, \vf^\prime, 
	{\vec D}\vf, \tau\right)
\end{equation}
governing the evolution between times $\tau_i$ and $\tau_f$.
The explicit time dependence in the action enters through factors of the metric, which appear in the covariant
volume element and terms with contracted derivatives. Computing
the variation of the action, subject to equations of motion and a fixed boundary condition
at $\tau_i$, we find, as in \eqref{svar}:
\begin{equation}
\label{momentumdensitydef}
 \frac{1}{\sqrt{\tg}}\, \frac{\delta S}{\delta \vf} 
	 =    \frac{\partial \Lag}{\partial \left( \frac{\vf'}{a} \right)} 
	 =  \pi \left(\vec{x},\tau\right)
	\end{equation}
where functional derivatives
are introduced with respect to coordinate volumes as in:
\begin{equation}
\delta S = \int \dd^3 x {\delta S\over \delta\phi}\delta\phi
\end{equation}

Computing the total time derivative of \eqref{firstscalaraction} we find the 
field theory version of the Hamilton-Jacobi equation
\begin{equation}
\label{firstHJ}
   H\left(\vf,\frac{1}{\sqrt{\tg}} \, \frac{\delta S}{\delta \vf}\,,\tau \right) 
	+ {1\over a(\tau)}~{\partial S \left(\vf,\tau\right)\over \partial \tau}  =  0
\end{equation}
This is a first order differential equation for the classical action as a functional of the boundary data 
for $\vf$. It is convenient for our purposes to introduce the densities:
\begin{equation}
\label{Sdensity}
S(\phi,\tau) = \int \dd^3 x\sqrt{\tilde g} ~\mathcal{S}(\phi,\tau)
\end{equation}
and:
\begin{equation}
\label{Hdensitya}
H(\phi,\pi,\tau) = \int \dd^3 x\sqrt{\tilde g}~\Ham (\vf,\pi,\tau) =  \int \dd^3 x\sqrt{\tilde g}~ 
(\pi \dot{\vf} - \Lag )
\end{equation}
Then the H-J equation becomes:
\begin{eqnarray}\label{densityHJ}
   \sqrt{g} \, \Ham\left(\vf, \frac{1}{\sqrt{\tg}}\, \frac{\delta S}{\delta \vf}, \tau \right) +  
   \frac{\partial}{\partial \tau} \left( \, \sqrt{\tg} \, \mathcal{S} \left(\vf, \tau \right) \,\right) & = & 0
\end{eqnarray}
In this form the equation is understood to hold up to total spatial derivatives.

\section{Infrared Divergences and their Interpretation}
\label{sec:strategy}
In this section we consider some simple examples of the infra-red divergences in inflationary 
spacetimes and discuss their interpretation. 

\subsection{The Free Scalar Field in de~Sitter Space}
\label{subsec:explicit}
It is instructive to begin the discussion with the simplest possible example: a
free scalar field evolving homogeneously in a fixed de~Sitter background.
In this case a straightforward way to compute the H-J functional explicitly is to integrate
the action by parts:
\begin{eqnarray}\label{firsscalaraction}\nonumber
	S & = & \int_{\tau_i}^{\tau_f} \bns \dd\tau\,\dd^3 x ~a^4 ~
	 \frac{1}{2} \left(  \, ( \frac{\vf\,'}
			{a})^2 - m^2\vf^2 \right) \\ 
			 &=&  \int d^3 x \left. {1\over 2}a^2 \vf \vf^\prime \right|^{\tau_f}_{\tau_i}-
			\int_{\tau_i}^{\tau_f} \bns \dd\tau\,\dd^3 x {1\over 2}\vf \left[ \partial_\tau
			(a^2\partial_\tau \vf) + a^4 m^2\vf	\right]	\nonumber \\ 				
			&=&  \int d^3 x \left. {1\over 2}a^2\vf \vf^\prime \right|^{\tau_f}_{\tau_i}
			\label{onsact}
\end{eqnarray}
The bulk term appearing in the intermediate step vanished because it is proportional
to the equation of motion, which is enforced when computing the
H-J functional. 

The H-J functional is supposed to depend  on the field, but not on its time derivative. The
final expression in \eqref{firsscalaraction} is therefore still not what we want. To
proceed we need the time dependence of the field. The metric of de~Sitter space 
is\footnote{The exponentially expanding coordinate system often used in cosmology
is recovered by the substitution $\tau=-e^{-Ht}$. Note that the conformal time
$\tau\in(-\infty,0)$ is negative. Also recall that these coordinate systems cover only 
half of de~Sitter space; 
%the past horizon $\tau\to -\infty$ can be reached in finite affine time.
$\tau \to -\infty$ corresponds to the past horizon.
}:
\begin{equation}
\label{frwmetric}
ds^2 = a(\tau)^2(-\dd\tau^2 + \dd \vec{x}\cdot \dd \vec{x})
\end{equation}
The de Sitter scale factor is $a(\tau) = -{1\over H\tau}$, with $H$ the (constant) Hubble parameter. The Klein-Gordon equation then becomes:
\begin{equation}
\label{scalareom}
	\vf\,'' - {2\over\tau} \,\vf\,' + {m^2\over H^2\tau^2} ~\vf  =  0
\end{equation}
This has the general solution:
\begin{eqnarray}
	\vf(\tau) & = & c_- \, \tau^{\lambda_{-}} + c_+ \, \tau^{\lambda_{+}} 
	\label{solsvf} \\
	\lambda_{\pm} & = & \frac{3}{2} \pm \frac{3}{2} \, \sqrt{1 - \left( {2m\over 3H}\right)^2}
	\label{lamdef}
	\end{eqnarray}
In this paper we only consider the case of a light field with 
%$m^2=\partial^2_\vf V|_{\vf=0}<9H^2/4$ 
$m^2 < 9 H^2 / 4$
so that the square root is real. 

Using the explicit solutions for $\vf$ we can now write the on-shell action \eqref{onsact} as:
\begin{equation} 
S = \left.  {1\over 2H^2} \int d^3 x ~\tau^{-3}~
(c_- \tau^{\lambda_-} + c_+ \tau^{\lambda_+})
(c_- \lambda_-\tau^{\lambda_-} + c_+\lambda_+\tau^{\lambda_+})
\right|^{\tau_f}_{\tau_i}
\end{equation}
To understand this expression, recall that the H-J functional is defined with fixed boundary 
conditions at some early time $\tau_i$. We have not yet specified these precisely. 
In the present context the natural choice is to take $\tau_i\to -\infty$ and impose regularity 
there. Physically this puts the field in its ground state.
Since $\lambda_+>3/2$ the terms in the action coming from the $\lambda_+$-solution diverge 
in the limit $\tau_i\to -\infty$. Our boundary condition therefore amount to taking $c_2=0$ 
and concentrating on the $\lambda_-$-branch. The limit $\tau_i\to -\infty$ 
gives no contribution for $\lambda_-$. The final result for the H-J functional 
becomes simply:
\begin{equation}
\label{sfinal}
S(\phi,\tau) \, = \, {1\over 2H^2}\int d^3 x ~\tau^{-3}c_-^{2}\lambda_- \tau^{2\lambda_-} \, 
= \, -{1\over 2}H \int d^3 x ~a^3(\tau)~\lambda_- \vf^2
\end{equation}

At this point we have computed the H-J functional in the simplest case. This allows
us to exhibit our first example of infra-red divergences, as follows. The integrand of the 
H-J functional 
\eqref{sfinal} scales as:
\begin{equation}
\label{firstdiv}
a^3 \vf^2 \sim \tau^{-3+2\lambda_-}=\tau^{-2\nu}
\end{equation}
where:
\begin{equation}
\label{nudef}
\nu = {3\over 2}\sqrt{ 1- \left( {2m\over 3H} \right)^2}
\end{equation}
is positive. Since late times correspond to $\tau\to 0$ the expression \eqref{firstdiv} diverges as the system
evolves to the asymptotic future. It is this type of divergence that we are interested in.

\subsection{Local Divergences and the Hamilton-Jacobi Functional}
\label{subsec:HJscalar}
Before discussing the interpretation of divergences, let us determine their 
form in a more general setting, using the Hamilton-Jacobi equation:
\begin{eqnarray}
\label{HJagain}
	\frac{1}{2} \, \left( \frac{1}{\sqrt{\tg}} \, \frac{\delta S}{\delta \vf} \right)^2 + \frac{1}{2} \, \vec{D} \vf \cdot  \vec{D} 
		\vf + V(\vf) + \frac{1}{\sqrt{g}} \,\partial_{\,\tau} \left( \sqrt{\tg} \,\mathcal{S}\right) & = & 0 
\end{eqnarray}
A large class of solutions to this equation are well approximated the {\it ansatz} :
\begin{eqnarray}
\label{Gexpansion}
	S = \int d^3 x \sqrt{\tilde{g}} ~\mathcal{S} & = & \int d^3 x \sqrt{\tilde{g}} ~\left[
	U(\vf) + M(\vf) \vec{D} \vf \cdot \vec{D} \vf + \ldots \right]
\end{eqnarray}
The effective expansion parameter in the derivative expansion  is the inverse metric.  
Since the inverse metric for de~Sitter space is $g^{ij}=(H \tau)^{2}\delta_{ij}$ 
we expect this type of expansion to be accurate at late times when $\tau$ is small. 
The same sort of ansatz has been used in studying holographic RG flows in the AdS/CFT 
correspondence \cite{deBoer:1999xf}, and a similar approach based on an expansion
 in spatial gradients was applied to inflationary
spacetimes in \cite{Salopek:1990jq, Salopek:1994sq, Salopek:1997he, Parry:mw}.
The {\it ansatz} \eqref{Gexpansion} gives the momentum density:
\begin{equation}
\label{HJmomentum}
	\pi= {1\over\sqrt{\tilde{g}}}~{\delta S\over \delta\vf} =  
	\partial_{\vf} U - \partial_{\vf}M \, \vec{D} \vf \cdot \vec{D} \vf - 2 M \, \vec{D}^2 \vf +\ldots
\end{equation}
In computing this expression we have discarded total \emph{spatial} derivatives that arise in the functional
derivative of \eqref{Gexpansion}.
Inserting the momentum in the H-J equation \eqref{HJagain} we find:
\begin{eqnarray}\label{FullscalarHJ}
	\frac{1}{2} \, \left( \partial_{\vf} U\right)^2 - \partial_{\vf} U \, \left( \partial_{\vf} M \vec{D} \vf \cdot \vec{D} \vf 
	+2 M \, \vec{D}^2 \vf \right)  + \frac{1}{2} \, \vec{D} \vf \cdot  \vec{D} \vf & &\\ \nonumber
	+ \, V(\vf) +  H \, \left(3 \,U(\vf) + M(\vf) \vec{D} \vf \cdot \vec{D} \vf \right)  +\ldots & = & 0
\end{eqnarray}
where the dots $\ldots$ again denote terms with more than two derivatives.
The last term in \eqref{FullscalarHJ} comes from the partial time derivative in the H-J equation \eqref{HJagain}. The
partial derivative only applies to the explicit time dependence due to factors of the metric, and not the implicit time
dependence of $\vf$.
Recalling that the H-J equation is valid only up to total spatial derivatives we solve 
\eqref{FullscalarHJ} order by order and find:
\begin{eqnarray} 
\label{Ueqn}
	\frac{1}{2}\, \left(\partial_{\vf} U \right)^2 + V(\vf) + 3 \,H\, U(\vf) & = & 0 \\ \label{Meqn}
	\frac{1}{2} + H M + 2 M \partial_{\vf}^{\,\,2} U + \partial_{\vf} M \, \partial_{\vf} U & = & 0
\end{eqnarray}
For a given theory, with a specific potential $V(\vf)$, the first equation
determines $U(\vf)$, and then the second equation yields $M(\vf)$. Since the first equation is nonlinear
it may in general be difficult to find a simple expression for $U(\vf)$.
However, we can always expand a regular potential $V(\vf)$ as a series in $\vf$ and then use \eqref{Ueqn} to
determine recursion relations for the coefficients in a corresponding expansion for $U(\vf)$.

Rather than pursuing this strategy generally, 
it is instructive to compute just the first few terms of such an expansion. We therefore consider a potential
with the leading term:
\begin{equation}
	V(\vf) = \frac{1}{2}  m^2  \vf^2 +\ldots
\end{equation}  
We have omitted a linear term, which can be cancelled by a redefinition of the field by an appropriate additive
constant. In addition, we have ignored the possibility of a constant term in the potential, which would not enter
the scalar equation of motion.
Solving \eqref{Ueqn} now gives:
\begin{equation}
\label{Usolutiona}
U(\vf)  = - \frac{1}{2} H  \lambda_- \vf^2+ \cdots
\end{equation}
More precisely, \eqref{Ueqn} allows for solutions with either of the $\lambda_\pm$
defined in \eqref{lamdef}; but we know from the discussion in the previous subsection that 
imposing regularity at early times corresponds to the $\lambda_-$ solution.
The expression \eqref{Usolutiona} agrees precisely with the result \eqref{sfinal}
found by explicit computation. However, the present computation is much more general 
because it shows that \eqref{Usolutiona} is also the correct leading term when considering 
spatially varying fields, or potentials with interactions. The important point here is that, instead
of finding the action as a function of $\tau$ as determined by some
particular solution, we have obtained the functional dependence on the field $\vf$.

Using the solution for $U(\vf)$ we can now solve \eqref{Meqn}, which gives:
\begin{equation} 
	M(\vf)  =  - \frac{1}{2 H \,\left( 1 - 2 \lambda_- \right)}  + \ldots
	  \label{Msolutiona}
\end{equation}
Since \eqref{Meqn} is a 
linear differential equation for $M$ we can freely add a multiple of the homogeneous solution 
\begin{equation}
M_{\rm hom}(\vf)  =  \vf^{\frac{1}{\lambda_-} - 2}
\end{equation}
to the particular solution \eqref{Msolutiona}. In inflationary scenarios we are generally interested
in a ``slowly rolling" scalar field whose kinetic energy is negligible compared to its potential
energy. This implies a very small mass $m^2 \ll H^2$, which in turn implies that $\lambda_- \ll 1$.
In this case, the homogenous solution $M_{\rm hom}(\vf)$ is of higher order than the terms we have
retained in the $\vf$-expansion, and so it is negligible. However, the homogenous solution might play
an important role in applications for which the slow roll condition does not apply.

Under the slow roll condition $\lambda_-\ll 1$ the field $\vf\sim\tau^{\lambda_-}$
depends only weakly on conformal time $\tau$. The scaling with $\tau$
of the terms in the derivative expansion \eqref{Gexpansion} is therefore dominated by 
factors of the metric. It follows that the approximate scalings are:
\begin{eqnarray}
	\sqrt{\tg} \, U(\vf) & \sim & \tau^{-3} \\
	\sqrt{\tg} \, M(\vf) g^{ij} & \sim & \tau^{-1} 
\end{eqnarray}
The subleading terms in the expansion, denoted by dots $\ldots$ in \eqref{Gexpansion}, 
all scale with positive powers~\footnote{Relaxing the slow roll condition to allow 
$\lambda_-\sim 1$ gives
faster convergence. A negative $m^2$ could give a more interesting divergence
structure; indeed, since this case is unstable it is expected that terms of higher order 
in the field play an important role.} of $\tau$. This means that $U(\vf)$ and $M(\vf)$, as determined by
equations \eqref{Ueqn} and \eqref{Meqn}, characterize \emph{all} divergences of the 
classical action.~\footnote{More precisely they compute all {\it local} 
divergences. These are power law divergences. We will consider logarithmic, or non-local,
divergences in due course.}

\subsection{Interpretation of Divergences}
In the previous section we demonstrated, through a simple example, that the H-J functional is
divergent at late times. Because the H-J functional is an on-shell action, we can try to interpret it
as we would an effective action in quantum field theory. In that context we are very familiar with
the appearance of ultra-violet divergences and their treatment through regularization and renormalization.

In the present problem the divergences appear at small conformal time $\tau$. However,
in de Sitter space, spatially inhomogenous waves depend on the dimensionless quantity
$k\tau$, so small $\tau$  is equivalent to small wave number $k$. The divergences 
we have found are
therefore large distance, or infra-red, divergences. Another way of seeing this is to 
consider how proper distances
change under a constant rescaling of the conformal time: $\tau \to \lambda \tau$. The de 
Sitter line element is:
\begin{eqnarray}
	\dd s^2 & = & \left(\,\frac{1}{H \tau}\,\right)^{2} \, \left( - \dd \tau^2 + \dd \vec{x} \cdot \dd \vec{x} \right)
\end{eqnarray}
Under the constant rescaling of $\tau$ this becomes:
\begin{eqnarray}
	\dd s^2 & \to & \left(\,\frac{1}{H \tau}\, \right)^{2} \,  \left( - \dd \tau^2 + \frac{1}{\lambda^2}\dd \vec{x} \cdot \dd \vec{x} \right)
\end{eqnarray}From the point of view of the metric, the same effect 
could be achieved by restricting oneself to a hypersurface
at a fixed $\tau$ and rescaling all of the spatial coordinates by $\vec{x} \to \lambda^{-1} \vec{x}$. Under such a
rescaling the wave number $k$ scales as $k \to \lambda k$. Therefore, a rescaling of the conformal time by a factor $\lambda$
can alternately be thought of as a rescaling of wave numbers by the same factor $\lambda$, keeping $\tau$ fixed. Thus, small $\tau$
divergences are indeed infra-red divergences. 

According to some interpretations of de~Sitter holography \cite{Strominger:2001pn, Maldacena:2002vr} there exists a dual description of the system
considered  here in which the infrared divergences are in fact ultraviolet divergences of a conventional
quantum field theory. We will not need to assume that a holographic interpretation of this sort exists because it seems clear
that, even if it does not, it is natural to deal with infrared divergences they way we normally treat ultraviolet
divergences. That is, we adopt a regularization scheme, introduce counterterms, and then renormalize.

The most straightforward way of regulating the divergences we have encountered is by simply `cutting spacetime off'
near the boundary. In de Sitter space we cut the space off at $\tau=\tau_0$, with $\tau_0$ a small negative number
\footnote{In our conventions $\tau$ is a negative number. The cut-off $\tau_0$ is a small negative number, and it is
understood that limits such as $\tau_0 \rightarrow 0$ are from below.}. Actions are then written as integrals 
over the regulated spacetime, which we denote $\MM_0$:
\begin{eqnarray}\label{scalaraction}
	S  =  -\int_{\MM_0} \bns \dd^4 x \, \sqrt{g} \, \left( \frac{1}{2} \, 
	\nabla_{\mu} \vf \, \nabla^{\mu} \vf + V(\vf) \right) 
\end{eqnarray}
When we refer to `the boundary' in calculations we mean $\dM_0$, though it is implied that at the end of a calculation 
the cutoff should be removed by taking $\tau_0 \rightarrow 0$. 

In the previous section we showed that all local infra-red divergences take the form
indicated by the two terms in \eqref{Gexpansion}. We can therefore cancel the divergences
by adding the counter-terms:
\begin{eqnarray}
\label{counterscalar}
	S_{\rm ct} & = & - \int_{\dM_0} \bns \dd^3 x \sqrt{\tg} \, \left( U(\vf) + 
	M(\vf) \, \vec{D} \vf \cdot \vec{D} \vf\right) 
\end{eqnarray}
to the action for the scalar field. The $U(\vf)$ and $M(\vf)$ take the functional form determined 
by \eqref{Ueqn} and \eqref{Meqn}. To the leading order they were given in \eqref{Usolution}
and \eqref{Msolution}. The renormalized action is the total action:
\begin{eqnarray}\label{totalscalaraction}
	S_{\rm tot} & = & S + S_{\rm ct}
\end{eqnarray}
It is in this expression that the cutoff can be removed by taking $\tau_0\to 0$.

The introduction of counterterms changes the action and it must be justified why
this is acceptable. One observation is that the counterterms only involve quantities 
intrinsic to the boundary. Adding such terms to the action does not change the bulk 
equations of motion. It therefore leaves bulk physics invariant, while making the action 
well-defined, even on a noncompact spacetime. The counterterms we have added are 
analogous the boundary counterterms that often appear in the AdS/CFT literature 
\cite{Balasubramanian:1999re,Emparan:1999pm,Kraus:1999di}.

Although sound, this reasoning does not fit with our interpretation of the action as a 
H-J functional. In this context the dependence on the boundary values obviously 
matters, it is all there is. The point here is that the infrared divergences are universal,
they take the same functional form for many different backgrounds. After subtracting
the divergences, the renormalized action $S_{\rm tot}(\vf)$ still depends on $\vf$, and 
this dependence is meaningful. The strategy is similar to that pursued in Pauli-Villars 
regularization of UV-divergences: simply subtract the action of a very massive field;
then that field will cancel the divergences, but leave a meaningful dependence
on the low-energy parameters. In the present construction the counterterms cancel 
the divergences by subtracting the action computed on a definite background action.
This will render finite and physically meaningful the effective action of fluctuations 
around this background. We will compute this action in the next section. 

In the present discussion of divergences we have assumed for simplicity that 
the background spacetime is de~Sitter. However, the approach is not limited
to de~Sitter space, or even spacetimes that are asymptotically de~Sitter.
The discussion applies in situations where de~Sitter space constitutes
a legitimate infra-red completion, {\it i.e.} the late time behavior can be chosen 
as de~Sitter. The actual late time behavior does not have to be de~Sitter space, 
anymore than a specific ultraviolet completion of a low energy, renormalizable 
field theory has to be taken seriously at arbitrarily high energies.

\section{Fluctuations and the Power Spectrum} 
\label{sec:fluctuations}

In this section we compute the Hamilton-Jacobi functional in a perturbation series away
from the homogenous solution. We find logarithmic divergences and interpret them
in terms of the power spectrum of the density fluctuations. 

\subsection{Introduction}
\label{subsec:flucintro}
We are interested in scalar fields in de~Sitter space because their
quantum fluctuations generate density perturbations which may have seeded the 
observed structure in the universe. Such fluctuations evidently have nontrivial spatial
dependence, in contrast to the homogenous solutions considered in the previous section.
The strategy for computing the H-J functional in this more general case is to treat fluctuations as
perturbations around a background homogeneous solution.
Thus we expand the action as:
\begin{eqnarray}
	S_{\rm tot}[\dvf] & = & S_{\rm tot}^{(0)} + S_{\rm tot}^{(1)} + \frac{1}{2} \, S_{\rm tot}^{(2)} + \ldots 
\end{eqnarray}
where $S_{\rm tot}^{(n)}$ represents the $n^{th}$ order variation of $S_{\rm tot}$, {\it i.e.}
it consists of terms with $n$ factors of the fluctuation $\dvf$. At each order
there are contributions from the action \eqref{scalaraction}, evaluated on-shell 
and with appropriate boundary conditions imposed, and there are also contributions from
the counterterms \eqref{counterscalar}. Thus we write:
\begin{eqnarray}
	S_{\rm tot}^{(n)} & = & \delta^n S \big\vert_{\vf } + \delta^n S_{\rm ct} \big\vert_{\vf }
\end{eqnarray}
where $\vf$ denotes the homogeneous background solution. 

The zeroth order term in this expansion vanishes:
\begin{eqnarray}
	S_{\rm tot}^{(0)} & = & S(\vf)- \int_{\dM_0} \bns \nts \dd^3 x \, \sqrt{\tg} \, \left( U(\vf) + M(\vf) \, 
			   	  \vec{D} \vf \cdot \vec{D} \vf  \right)				 = 0
\label{zerothorder}
\end{eqnarray}
Indeed we computed each of these terms in section \ref{sec:strategy}: in \ref{subsec:explicit} 
we computed the on-shell action for the explicit homogeneous solution, 
and in \ref{subsec:HJscalar} we
solved the H-J equation to find the local part of the H-J functional. The two results agreed.
Since the zeroth order term \eqref{zerothorder} is the difference between these, it vanishes.
Of course, counter-terms were chosen to make this happen. 

Varying $S$ in \eqref{counterscalar} to obtain the first order terms we find:
\begin{eqnarray}
	S^{(1)} & = & \int_{\MM_0} \bns \dd^4 x \, \sqrt{g} ~ \dvf \left(  \nabla^{2} \vf - \partial_{\vf} V \right) +
		\int_{\dM_0} \bns \nts \dd^3 x \, \sqrt{\tg} \, ~\dvf \, \frac{\vf\,'}{a}
		\label{firstvar}
\end{eqnarray}
after integration by parts. The bulk term vanishes because we impose the equation of motion. 
This leaves only the boundary term in \eqref{firstvar}. Similarly varying the counter-term
\eqref{counterscalar} gives:
\begin{eqnarray}
\label{firstctvar}
	S_{\rm ct}^{(1)} & =  & - \int_{\dM_0} \bns \dd^3 x \, 
	\sqrt{\tg} \, \left( \dvf \, \partial_{\vf} U + \dvf \, \partial_{\vf} M \, 
				 \vec{D} \vf \cdot \vec{D} \vf  + 2 M \, \vec{D} \dvf \cdot \vec{D} \vf \right) \\
			    & = & - \int_{\dM_0} \bns \dd^3 x \, \sqrt{\tg} \, \left( \partial_{\vf} U - \partial_{\vf} M \, 
				 \vec{D} \vf \cdot \vec{D} \vf  - 2 M \,  \vec{D}^2 \vf \right) \, \dvf
\end{eqnarray}
Adding the two equations, and referring back to the two expressions for the momentum
given in equations \eqref{simplemom} and \eqref{HJmomentum}, we see that
the first order term in the action vanishes on-shell:
\begin{eqnarray}
	S_{\rm tot}^{(1)} & = & 0
	\label{stotone}
\end{eqnarray}
The equality verified here is of course nothing but the general relation \eqref{momentumdensitydef}.

The vanishing of the first order variation around a solution of the equations of motion is precisely the condition that 
the solution extremizes the action. The result \eqref{stotone} is therefore hardly surprising.
It should be noted, however, that the computation here is distinct from the usual result in 
classical 
field theory. The standard variational principle involves fixing $\vf$ on the boundary so that
$\dvf = 0$ automatically on the boundary. Here we are considering a finite portion of de Sitter 
space, with \emph{arbitrary} Dirichlet boundary conditions on the scalar field at the boundary $\tau=\tau_0$. In
other words, the first order variation of the total action vanishes on-shell, despite the fact that $\dvf \neq 0$
on the boundary. The contributions from the boundary counterterms precisely cancel the terms that we would
normally discard by requiring $\dvf=0$ on the boundary.

\subsection{Quadractic Fluctuations}
The zeroth and first order terms in our expansion of the action both vanish. The first non-zero
contribution to the action comes from terms quadratic in $\dvf$, which lead to density perturbations.

There is a trick to compute the variation of the action to the second order:
note that the first order variation \eqref{firstvar} is valid for all $\vf$, whether
they satisfy the background equation of motion or not. We can therefore determine the 
second variation by varying $\vf$ in  \eqref{firstvar}. This immidiately gives:
\begin{eqnarray}
\label{stwoeq}
	S^{(2)} & = & \int \dd^4 x \, \sqrt{g} \, \dvf \, \left(\nabla^{2} \,  - \partial_{\vf}^{\,\,2} V \right) \dvf
				+ \int_{\dM_0} \bns \dd^3 x \, \sqrt{\tg} \, \frac{1}{a} \,  \dvf \, \dvf\,' 
\end{eqnarray}
The condition for the bulk piece to vanish is the equation of motion for $\dvf$:
\begin{eqnarray}\label{dvfeom}
	\dvf\,'' + 2 \, \HH \, \dvf \, ' - a^2\vec{D}^2 \, \dvf + a^2 \, \partial_{\vf}^{\,\,2} V \, \dvf & = & 0
\end{eqnarray}
Here $\HH=a^\prime/a$ is related to the Hubble expansion factor by $H=\HH/a$.

After imposing the equations of motion, the quadratic action reduces to the boundary 
integral in \eqref{stwoeq}. This integral must be treated with care because of the interplay
between the time derivative and our choice of regularization procedure. The precise meaning
 of the integral is:
\begin{eqnarray}
\label{normalderivativeterm}
S^{(2)} =   \int_{\dM_0}  \bns\nts \dd^3 x \, a(\tau_0)^2 \, \dvf(\vec{x},\tau_0) \, \lim_{\tau \rightarrow \tau_0} 
\frac{\partial \,\dvf(\vec{x},\tau)}{\partial \tau}
\end{eqnarray}
It is convenient to work with the Fourier transform of the fluctuation $\dvf$ and 
so write:
\begin{eqnarray}
\label{stwodef}
S^{(2)} & = & \int \dd^3 k~ \dd^3 p~ \delta^{(3)}(\vec{k}+\vec{p})\, \dvf_{\vec{p}}(\tau_0) \, \dvf_{\vec{k}}(\tau_0) 
\, F_{\vec{k}}(\tau_0)
\end{eqnarray}
where:
\begin{eqnarray}\label{Fdefn}
	F_{\vec{k}}(\tau_0) & = & a(\tau_0)^2 \, \lim_{\tau \rightarrow \tau_0} 
		\frac{\partial}{\partial \tau} \left( \frac{\dvf_{\vec{k}}(\tau)}{\dvf_{\vec{k}}(\tau_0)} \right)
\end{eqnarray}
Re-expressing the normal derivative of a field in terms of its boundary data at the cut-off is a familiar procedure
in AdS/CFT calculations, which are technically similar to the approach we take here. An excellent discussion of this
procedure and its physical meaning can be found in \cite{Freedman:1998tz}.

As in section \ref{subsec:explicit}, the evaluation of the boundary term ultimately 
requires the solution of the bulk equation of motion. The equation of 
motion \eqref{dvfeom} for $\dvf_{\vec{k}}(\tau)$ is:
\begin{eqnarray}
	\dvf_{\vec{k}}\,'' + 2 \,\HH \, \dvf_{\vec{k}}\,' + \left(k^2 + a(\tau)^2 m^2 \right) \, \dvf_{\vec{k}} & = & 0
\end{eqnarray}
In de Sitter space where $a(\tau)=-{1\over H\tau}$ this is essentially the
Bessel equation. Solutions are of the form:
\begin{eqnarray}\label{dvfsoln}
	\dvf_{\vec{k}}(\tau) & = & |\tau|^{3/2} \, \left( c_1 \, J_{-\nu}(|k \tau|) + 
	c_2 \, J_{\nu}(|k \tau|)\right)
\end{eqnarray}
where $\nu$ was given in \eqref{nudef}. The regularity condition that the solution contains 
only a positive frequency component at $\tau \rightarrow - \infty$ determines the constants $c_1$ and 
$c_2$ up to a common factor. The classical solution then becomes:
\begin{eqnarray}\label{posfreqsoln}
	\dvf_{\vec{k}}(\tau) & = & |\tau|^{3/2} \, H \, \sqrt{\frac{\pi}{4}} \, \left(  J_{-\nu}(|k \tau|) - 
	e^{ \pi i \nu} \, J_{\nu}(|k \tau|)\right)
\end{eqnarray}
The overall normalization is not needed in our approach but, for definiteness,
is determined up to an overall phase using:
\begin{eqnarray}\label{kgnormcond}
	c_1 \, c_{2}^{*} - c_2 \, c_{1}^{*} &  = & \frac{i \pi}{2 \sin\left(\pi\nu\right)} \, H^2
\end{eqnarray} 
which follows from the Klein-Gordon normalization condition on the modes 
$\dvf_{\vec{k}}$. 

Using these modes we can now  evaluate the function $F_{\vec{k}}$ from \eqref{Fdefn}
and then the H-J functional from \eqref{stwodef}. Due to the Bessel functions the general 
result is quite messy and not illuminating. Expanding the result in the small
parameter $\tau_0$ gives the more manageable expression:
\begin{eqnarray}\label{finalF}
	F_{\vec{k}}(\tau_0) & = & \frac{\lambda_-}{H^2 \, \tau^3_0} +
	\frac{k^2 \, }{H^2 (1-2 \lambda_-)\, \tau_0} +\frac{i k^3}{H^2}(k\tau_0)^{-2\lambda_-}	+ {\cal O}( (k\tau_0)^{2-2\lambda_-},(k\tau_0)^{3-4\lambda_-})
\end{eqnarray} 
where, as in earlier computations, $\lambda_-={3\over 2}-\nu$. The scaling of the leading
correction is one of the terms indicated, depending on the value of $\lambda_-$.
For small or modest $\lambda_-$ either correction vanishes when the cutoff is 
removed by taking $\tau_0 \rightarrow 0$. We are primarily interested in the slow-roll 
case where $\eta = \frac{m^2}{3 H^2}  \ll  1$ so, indeed, $\lambda_-\ll 1$. 
In fact, we have already used the slow roll condition to simplify the
otherwise complicated coefficient of $(k\tau_0)^{-2\lambda_-}$ in \eqref{finalF}. 

At this point we have computed the full H-J functional to quadratic order, 
using the explicit solutions to the equation of motion. Let us now consider the 
counterterms. The simplest way to compute the second variation of the counterterm
\eqref{counterscalar} is to note, again, that first variations such as \eqref{firstctvar}
are valid for all $\vf$, whether they satisfy the equations of motion or not. Thus
we can simply vary again and find:
\begin{eqnarray} 
\label{sttt}
	S_{\rm ct}^{(2)} 
	& = & - \int_{\dM_0} \bns \dd^3 x \, \sqrt{\tg} \, \left( \dvf^2 \, \partial_{\vf}^{\,\,2} U + \dvf^2 \, 
		\partial_{\vf}^{\,\,2} M \, \vec{D} \vf \cdot \vec{D} \vf  + 4 \, \dvf \, \partial_{\vf} M \, 
		\vec{D} \dvf \cdot \vec{D} \vf  \right. \\ \nonumber
		& & \;\;\;\;\;\;\;\;\;\;\;\;\; \left. + 2 \, M \, \vec{D} \dvf \cdot \vec{D} \dvf \right) 
\end{eqnarray}
This is the general result. The expression simplifies in the present context because
we are considering a spatially homogeneous background field $\vf$ and so 
$\vec{D} \vf=0$. 

In the case where the potential is dominated by a simple mass term we computed the
functions $U(\vf)$ and $M(\vf)$ explicitly in section \ref{subsec:HJscalar}, with the
results given in \eqref{Usolutiona} and \eqref{Msolutiona}. Inserting these expressions in
\eqref{sttt} and decomposing the fluctuations into Fourier modes $\dvf_{\vec{k}}$ gives:
\begin{eqnarray}
\label{almostthere}
	S_{\rm ct }^{(2)}  & = & 
	\int \dd^3 k~\dd^3 p \, ~\delta^{(3)}(\vec{k}+\vec{p}\,) \,\, a(\tau_0)^3  
	\, \left(  H \, \lambda_- -
	 	\frac{\vec{k} \cdot \vec{p}}{H\, (1-2\lambda_-)a(\tau_0)^2}   \right)
		\dvf_{\vec{p}} \, \dvf_{\vec{k}}
\end{eqnarray}

The total quadratic action is the sum of the ``bare" H-J functional \eqref{stwodef}, using the 
function $F_{\vec{p}}$ given in \eqref{finalF}, and the counterterm \eqref{almostthere}.
We find the  total action
\begin{eqnarray}\label{finalscalaraction}
	S_{\rm tot} [\dvf_{\vec{k}}\,] & = & 
	\int \dd^3 k \, \dd^3 p \,\, \delta^{(3)}(\vec{k}+\vec{p}\,) \, \frac{i k^3}{2 H^2} \, 
			\left( k \tau_0 \right)^{-2 \lambda_-} \, \dvf_{\vec{k}} \,  \dvf_{\vec{p}}
\end{eqnarray}
The contributions from the counterterms have completely removed the divergences
appearing in the first two terms of \eqref{finalF}. Referring to the computation of $F_{\vec{k}}$, it
is clear that those terms are due to the first part of the mode \eqref{dvfsoln}, which is proportional
to $\tau^{3/2} J_{-\nu}(k \tau)$. This part of the mode, which is small at early times, is dominant near
the boundary $\tau=\tau_0$. It is therefore reasonable that the counterterms obtained from 
the H-J
equation completely cancel the power-law divergences. The third term in $F_{\vec{k}}$ depends
crucially on the second part of the mode \eqref{dvfsoln}, proportional to $\tau^{3/2} J_{\nu}(k\tau)$.
This function plays a role in the regularity of the solution at early times but is small near the boundary. 
Because it is subleading compared to the first term in \eqref{dvfsoln}, this part of the mode is not
captured by the local arguments that determine the counterterms, and its contribution to the total
action survives unmodified. In this sense the quadratic action \eqref{finalscalaraction} is truly nonlocal.
\footnote{A precise statemens of the qualitative remarks in this paragraph is that the 
first two terms in \eqref{finalF} are independent of our choice of $c_1$ and $ c_2$, 
while the third term depends on the ratio $c_2 / c_1$.}

We now extract the power spectrum from the effective action \eqref{finalscalaraction},
following \cite{Maldacena:2002vr}. The semiclassical wavefunction is:
\begin{eqnarray}
	\Psi [\dvf] & \sim &  \exp\left( i S_{\rm tot}[\dvf] \right) 
\end{eqnarray}
and the two-point correlation function for the fluctuations $\dvf_{\vec{k}}$ is given by:
\begin{eqnarray}
	\left< \,\dvf_{\vec{k}} \,\dvf_{\vec{p}}\,\right> & = & \int \DD \dvf \, ~\dvf_{\vec{k}} \, \dvf_{\vec{p}} \,  \left| \Psi[\dvf ] \right|^2
\end{eqnarray}
After performing the Gaussian integral we find:
\begin{eqnarray}
\label{speccorr}
	\left< \,\dvf_{\vec{k}} \,\dvf_{-\vec{k}}\,\right> & = & \frac{H^2}{2  k^3} \, \left( k \tau_0 \right)^{2\eta} 
\end{eqnarray}
The power spectrum is related to the two-point correlator by:
\begin{eqnarray}
	P_{\dvf} (\vec{k}) & = & \frac{k^{3}}{2 \pi^2} \, \left< \,\dvf_{\vec{k}} \,\dvf_{-\vec{k}}\,\right>
\end{eqnarray}
This gives the power spectrum:
\begin{eqnarray}
\label{specspec}
	P_{\dvf}(\vec{k}) & = & \left( \frac{H}{2 \pi}\right)^2 \, \left( k\tau_0\right)^{2\eta}
\end{eqnarray}
For the massless scalar this reduces to the standard scale-invariant result. A small mass gives rise
to logarithmic corrections leading to mildly broken scale invariance. Note that, in these
last few equations, we use the notation $\eta=m^2/3H^2\simeq\lambda_-$ which
is conventional in cosmology.

\subsection{Comments on Logarithmic Divergences}
\label{logdivs}
As it stands, our final result \eqref{specspec} for the power spectrum depends on the cut-off $\tau_0$.
Although this dependence is just logarithmic, rather than a power-law, it is clearly 
not acceptable to have divergences, however mild, in physical quantities.

The interpretation of these divergences can be understood by inspecting the total 
action \eqref{finalscalaraction}. We have introduced $S_{\rm tot}$ as a renormalized
action but it appears to depend explicitly on $\tau_0$. However, from 
\eqref{posfreqsoln} we see that classical modes scale as 
$\dvf_{\vec{k}}\sim\tau_0^{\lambda^-}$ as $\tau_0\to 0$ so, in fact, the dependence 
on $\tau_0$ disappears as the cut-off is removed $\tau_0\to 0$. The action is
therefore truly renormalized. 

It is clear from this example that the $\tau_0$-dependence of the correlator \eqref{speccorr} 
simply reflects $\tau_0$-dependence of the fields. The cut-off $\tau_0$ acts 
like the renormalization scale that is well-known from UV renormalization theory. 
The total, renormalized, action does not depend on the scale, but several of
the objects it is written in terms of do. The scale dependence can be removed from 
physical observables but it appears in many of the quantites we define at intermediate steps 
of the computation.  

The discussion so far mimics the standard, somewhat formal, renormalization theory. 
A more direct way to get at the physics may be to simply interpret $\tau_0$ as a physical 
cut-off, along the lines of Wilson's approach to renormalization. Since we are considering 
infra-red divergences this amounts to choosing the cut-off $\tau_0$ as the {\it lowest} 
scale appearing in the problem. A reasonable choice in de~Sitter space would 
then be the de~Sitter scale $H$. Since the physical momentum is related to the 
coordinate momentum used in computations as $k_{\rm phys} = k/a$ this identification 
amounts to $\tau_0\sim 1/aH$. \footnote{It is amusing to 
contemplate the holographic interpretation of this prescription: we are led to introduce
an effective holographic screen at the time of horizon crossing.}

The main lesson of our computation is thus that the initial, apparently severe 
powerlaw divergences are in fact benign: a subtraction procedure can be devised 
that decouples physical quantities from the problems in the far infra-red. The theory
might not have behaved this way; it could have been that detailed assumptions
about the infrared would feed into physical quantities. 

The notorious fine-tuning problems of inflation, usually
thought of as arising in the UV, have some similarities with the issues addressed here.
Since the IR and UV behaviors are in fact related in
gravitating theories our  considerations may have some bearing on these problems. This
holds for dark energy as well, whose fine-tuning problems seem even more severe than
those of inflation.

Before concluding this section, let us point out an additional, conceptual, reason 
that we must introduce a cut-off: the asymptotic future ${\cal I}^+$ of de~Sitter space 
has the property that any two points are spacelike separated. Correlation functions 
therefore do not represent quantities that are measurable in the conventional sense 
although, perhaps, they could be afforded some sort of reality as ``meta-observables"~\cite{Witten:2001kn, Danielsson:2002qh}.
In the presence of a cut-off the asymptotic future does not have to be de~Sitter, so we can introduce a more conventional
inflationary spacetime where the correlators reenter the horizon and become
observable as the structure of the universe.

\section{The Gravity-Scalar System}\label{sec:gravscalar}
In previous sections we considered a scalar field propagating on a fixed background.
In this section we incorporate the back reaction on the spacetime and so consider
the combined scalar-gravity system. We construct counterterms using the H-J
formalism and show that they cancel all power-law divergences. We identify
the logarithmic divergences of quadratic fluctuations in slow-roll inflation and 
recover the scalar spectral index $n_s$. 
%known from realistic models of inflation. 

\subsection{Introduction}
\label{sec:GSslowroll}

The action for a scalar field coupled to gravity is:
\begin{eqnarray}\label{scalargravityaction}
	S & = & \int_{\MM_0} \nts \dd^4 x \sqrt{g} \, \left( \frac{1}{16 \pi G} \, R - \frac{1}{2} \, \nabla^{\mu} \vf \nabla_{\mu} \vf - V(\vf) \right)
		- \frac{1}{8 \pi G} \int_{\dM_0} \nts \dd^3 x \sqrt{\tg} \, K
\end{eqnarray}
The  Gibbons-Hawking boundary term, proportional to the trace of the extrinsic curvature, 
ensures that the action represents a well defined variational problem~\cite{Gibbons:1976ue}. 
It should not be confused with the boundary terms we add as counterterms. The latter are
formed from the intrinsic geometry of the boundary and have no bearing on the variational
principle.

Although the validity of our methods is more general than the examples given here, we will for
the most part consider spatially flat FRW cosmologies as backgrounds.
We will then study general fluctuations around this background to quadratic order.
The equations of motion for the background are thus the FRW equations:
\begin{eqnarray}
	\vf\,'' + 2 \HH \, \vf\,' + a^2 \pvf V & = & 0
\end{eqnarray}
\begin{eqnarray}
	\frac{3}{8\pi G} \, \left(\frac{\,\HH\,}{a}\right)^2 & = & \frac{1}{2} \, \left(\frac{\vf\,'}{a}\right) ^2 + V
	\label{frwtwo}
\end{eqnarray}
As before primes denote derivatives with respect to conformal time, and $\HH = \frac{\,a\,'}{a}$. 

As in previous sections it is essential that spacetime is effectively de~Sitter. For example,
this is needed for the H-J equation to determine the structure of divergences.
Although some of our results will be more general, we will mostly ensure this by
specializing the background to ``slow roll inflation", {\it i.e.} configurations with slowly 
evolving scalar fields. The energy density of the scalar field is then dominated by its potential
energy, which changes very slowly due to the small time derivatives of the field. In effect, the
potential energy of the scalar field acts like a cosmological constant. As is customary, we
define the slow-roll parameters $\eps$ and $\eta$:
\begin{eqnarray} \label{GSeps}
	\eps & = & \frac{1}{16 \pi G} \, \left( \frac{\partial_{\vf} V}{V} \right)^2 \\ \label{GSeta}
	\eta & = & \frac{1}{8\pi G} \, \frac{\partial_{\vf}^{\,2} V}{V} 
\end{eqnarray}
Slow-roll inflation corresponds to $\eps \ll 1$  and $\eta \ll 1$. We will frequently
work at linear order in the slow-roll parameters, making no assumptions 
about their relative magnitudes. 

When we are in the slow-roll regime the parameters $\eps$ and $\eta$ can be treated as constant with respect
to the conformal time. This can be seen by taking the derivative of $\eps$ or $\eta$ and using the equations
of motion to show that the resulting expression is quadratic in the slow-roll parameters and therefore negligible.
In our computations we will also need the following alternate expressions for $\eps$ and $\eta$:
\begin{eqnarray} \label{alteps1}
	 \eps & =  & 4 \pi G \, \left( \frac{\,\vf\,'}{\HH}\right)^2  =  1 - \frac{\,\HH\,'}{\HH^2} \\
	 & & \eta - \eps \, \,  = \, \,  1 - \frac{\,\vf\,''}{\HH \, \vf\,'}
\end{eqnarray}
These expressions follow from the equations of motion and are valid up to terms quadratic in the slow-roll parameters. 

\subsection{The Local Counterterms}
\label{sec:GSHJeqn}
We now compute the local counterterms for the combined scalar-gravity system. We will consider a general spacetime:
\begin{eqnarray}
\label{metricansatz}
	\dd s^2  =  g_{\mu\nu}\dd x^\mu  \dd x^\nu 
\end{eqnarray}
but it will be convenient to specialize the metric slightly from the outset,
by choosing a gauge with vanishing time-space component $g_{\tau i}=0$. This
is not mandatory but it simplifies the time+space split that is integral to the H-J 
formalism and natural in cosmological applications. 

With our choice of gauge we can write the action as:
\begin{eqnarray}\label{GCaction}
	S & = & \int_{\MM_0} \bns \dd^4 x \,\sqrt{g} \, \left( \frac{1}{16 \pi G} \, \left( \RR + K^{ij}K_{ij} - K^2\right) - 
		\frac{1}{2} \, \nabla^{\mu} \vf \nabla_{\mu} \vf - V(\vf) \right) 
\end{eqnarray}
where $\RR$ is the curvature of the three dimensional slice, and $K_{ij}$ is the extrinsic
curvature.
The computation leading to this result involves rewriting the four-dimensional curvature
using the Gauss-Codazzi equations (see \eqref{GCthree} in appendix \ref{app:Conventions}) 
and integrating by parts. Note that the Gibbons-Hawking boundary term cancelled. 

Taking the scalar $\vf$ and the spatial part of the metric $g_{ij}$ as the fundamental fields, the canonical momenta derived from \eqref{GCaction} are:
\begin{eqnarray}
	\pi_{\vf} & = &  {\vf^\prime\over\sqrt{-g_{00}}} \\
	\label{piijdef}
	\pi^{ij}  & = & \frac{1}{16 \pi G} \, \left( K^{ij} - g^{ij} K\right)
\end{eqnarray}
The Hamiltonian density for the gravity-scalar system is then:
\begin{eqnarray}
	\Ham & = & 16 \pi G \left( \pi^{ij} \pi_{ij} - \frac{1}{2} \, \pi^{i}_{\,i} \, \pi^{j}_{\,j} \right) + \frac{1}{2} \, \pi_{\vf}^{\,2}
		+ \frac{1}{2} \, \vec{D} \vf \cdot \vec{D} \vf + V(\vf) - \frac{1}{16 \pi G} \, \RR
\end{eqnarray}

As in the case of a fixed background, the H-J functional is the on-shell action, written
in terms of the fields evaluated at some late time. The Hamilton-Jacobi equation is
derived, again, by differentiating with respect to the time. This simply gives
the Hamiltonian constraint $\Ham=0$ which, in terms of the H-J functional
$S(\vf,g_{ij})$, reads:
\begin{eqnarray} \label{scalargravityHJ}
	16\pi G \left[ \frac{1}{2}\,\left(\frac{1}{\sqrt{\tg}}\,g_{ij} \frac{\delta S}{\delta g_{ij}}\right)^2 - 
		         \left(\frac{1}{\sqrt{\tg}}\, \frac{\delta S}{\delta g_{ij}}\right) 
			\left(\frac{1}{\sqrt{\tg}} \frac{\delta S}{\delta g^{ij}}\right)\right] - 
			\frac{1}{2} \left(\frac{1}{\sqrt{\tg}} \frac{\delta S}{\delta \vf}\right)^2
	 &  & \\ \nonumber
         = \, \, \, V - \frac{1}{16\pi G} \, \RR + \frac{1}{2} \, \vec{D} \vf \cdot \vec{D} \vf & &
\end{eqnarray}
Note that, in contrast to the H-J equation in section \ref{sec:HJTheory}, there is no
term $\partial_t S$ that takes explicit time dependence into account. 
This is because time translations are diffeomorphisms, and including the metric as a dynamical field removes
any explicit time dependence from the action. \footnote{Recall that, in the previous example of a scalar field on a fixed background spacetime, the
explicit time dependence in the action was entirely due to factors of the metric.}

Our interest in the H-J equation is, as in previous sections, that it allow us to isolate the 
local part of the H-J functional. Since the power-law divergences are contained entirely in the
local part of the H-J functional it will essentially be our counterterm. An appropriate local ${\it ansatz}$ for the Lagrangian,
expanded up to terms with two derivatives (one factor of the inverse metric), is now:
\begin{eqnarray}\label{sgclassicalaction}
	S & = & \frac{1}{8 \pi G} \, \int_{\dM} \nts \dd^3 x \sqrt{\tg} \, \left( U(\vf) + M(\vf) \vec{D} \vf \cdot \vec{D} \vf
			+ \Phi (\vf) \RR + \ldots \right)
\end{eqnarray}
The corresponding canonical momenta are:
\begin{eqnarray} \label{HJscalarmomentum} 
	\pi_{\vf} & = & \frac{1}{\sqrt{\tg}} \, \frac{\delta S}{\delta \vf} \\ \nonumber
			 & = & \frac{1}{8 \pi G} \, \left( \partial_{\vf} U - \partial_{\vf} M \, \vec{D} \vf \cdot \vec{D} \vf 
				- 2 M \, \vec{D}^2 \vf  + \RR \, \partial_{\vf} \Phi \right) \\ \label{HJmetricmomentum}
	\pi^{ij} & = & \frac{1}{\sqrt{\tg}} \, \frac{\delta S}{\delta g_{ij}} \\ \nonumber
			& = & \frac{1}{8 \pi G} \, \left( \frac{1}{2} g^{ij} \left( U + M \, \vec{D} \vf \cdot \vec{D} \vf \right) - M \, D^i \vf D^j \vf
					- \Phi \, \mathcal{G}^{ij} + D^{i} D^{j} \Phi - g^{ij} \, \vec{D}^2 \Phi \right)
\end{eqnarray}
In the last expression we used $\GG_{ij}$ to denote the Einstein tensor of the induced metric:
\begin{eqnarray}
	\GG_{ij} & = & \RR_{ij} - \frac{1}{2} \, g_{ij} \, \RR
\end{eqnarray}
We now evaluate the Hamilton-Jacobi equation \eqref{scalargravityHJ} using these expressions. Collecting functionally independent terms 
we obtain three equations:
%\footnote{In the previous example the two terms in the classical action were distinguished by their order in %an inverse metric expansion. In this example two of the three terms, proportional to $\vec{D} \vf \cdot %\vec{D} \vf$ and $\RR$, have the same number of inverse metrics, but are considered to be functionally %independent terms.} 
\begin{eqnarray}\label{GSUeqn}
	V + \frac{1}{2} \left( \frac{1}{8\pi G} \, \partial_{\vf} U\right)^2 - \frac{3}{32 \pi G} \, U^2 & = & 0
\end{eqnarray}
\begin{eqnarray}\label{GSPhieqn}
	\frac{1}{2} \, (1 + U \, \Phi ) - \frac{1}{8 \pi G} \, \partial_{\vf} U \, \partial_{\vf} \Phi & = & 0
\end{eqnarray}
\begin{eqnarray}\label{GSMeqn}
	\frac{1}{2} - \frac{1}{16 \pi G} \, U \, M - \frac{1}{4\pi G} \, \partial_{\vf}U \, \partial_{\vf} \Phi + 
		\left( \frac{1}{8 \pi G} \right)^2 \, \left( \partial_{\vf} U \, \partial_{\vf} M +2 M \partial_{\vf}^{\,2}U \right) & = & 0
\end{eqnarray}
These equations determine the functions $U(\vf)$, $M(\vf)$, and $\Phi(\vf)$ in the H-J 
functional \eqref{sgclassicalaction}. When evaluated on a quasi-de~Sitter background these
will be the only terms that diverge as the cut-off $\tau_0$ is taken to the asymptotic future.

The first equation \eqref{GSUeqn} is a non-linear differential equation for $U(\vf)$ which, in general,
is difficult to solve. For a specific potential $V(\vf)$, if we can solve for $U(\vf)$ it is then straightforward to
integrate the linear (albeit inhomogenous) equations \eqref{GSPhieqn} and \eqref{GSMeqn} to obtain $\Phi(\vf)$
and $M(\vf)$. As we saw in the case of a scalar field on a fixed background, one is free to supplement
the resulting expressions for $\Phi(\vf)$ and $M(\vf)$ with solutions of the corresponding homogenous equations.

Luckily, there is a nice trick for finding $U(\vf)$. Consider temporarily a flat FRW cosmology with 
scale factor $a(\tau)$ and a spatially homogeneous scalar field $\vf(\tau)$. For such a 
configuration 
the two equations \eqref{piijdef} and \eqref{HJmetricmomentum} both give simple
expressions for $\pi_{ij}$. Comparing the results we find:
\begin{equation}
 U(\vf)  = - 2 \,\frac{\HH}{a}
\label{Umaster} 
\end{equation}
in units where $8 \pi G = 1$.
The function $U(\vf)$ is therefore essentially the standard Hubble parameter 
$H=\dot{a}/a=\HH/a$, expressed in terms of the scalar field. $H(\vf)$ is often considered 
in cosmology\footnote{Indeed, in the literature the use of $H(\vf)$ is often referred to as 
the H-J formalism. The development of the subject was initiated in \cite{Salopek:1990jq} }, 
but its interpretation as the counterterm $U(\vf)$ seems new. It is important to emphasize that the assumption
of an FRW cosmology only plays an auxiliary role in obtaining this result. Once we have determined
the functional $U(\vf)$ we can use it for general backgrounds and scalar field configurations
that may be spatially dependent. The
universality of the local terms in the H-J functional is precisely what makes them suitable as counterterms.

Let us carry out this procedure in the slow-roll case. Combining the FRW equation
\eqref{frwtwo} and the expression \eqref{alteps1} for the slow roll parameter we compute
$\HH/a$ and then \eqref{Umaster} gives:
\begin{eqnarray} 
\label{Usolution}
	U(\vf) & = & -2 \sqrt{\frac{\,V(\vf)\,}{3 - \eps}}\simeq -2\sqrt{V\over 3}\left[1 + {1\over 12}
	\left({\partial_\vf V\over V}\right)^2 \right]
\end{eqnarray}
This is $U(\vf)$ for a general potential satisfying the slow roll conditions.

Next, we solve \eqref{GSPhieqn} to  find $\Phi(\vf)$. The {\it ansatz}:
\begin{eqnarray}
	\Phi(\vf) & = & \frac{f(\vf)}{U(\vf)}
\end{eqnarray}
gives:
\begin{eqnarray}\label{trialsolutioneq}
	1 + f + 2 f \, \left(\frac{\,\partial_{\vf} U\,}{U}\right)^2 - 2 \,\frac{\partial_{\vf} U}{U} \, \partial_{\vf} f & = & 0
\end{eqnarray}
As noted in section \ref{sec:GSslowroll}, the slow roll parameters are constants of
motion in the slow roll approximation up to terms of second order in slow roll. 
Differentiating \eqref{Usolution} and using the definition of $\eps$ we therefore find:
\begin{eqnarray}
	2 \left( \frac{\partial_{\vf} U}{U} \right)^2  & = &  \eps 
\end{eqnarray}
to the leading order. It is then clear that \eqref{trialsolutioneq} expresses $f$ in terms
of $\epsilon$ only, and so it is consistent to assume $f$ is a constant of motion 
as well. The remaining equation is then algebraic. The final result is:
\begin{eqnarray}
	\Phi(\vf) & = &  \frac{\eps-1}{\,U(\vf)\,}\simeq \sqrt{3\over 4V} \left[ 1 - {7\over 12}
	\left({\partial_\vf V\over V}\right)^2 \right]
	\label{Psolution}
\end{eqnarray}

A similar computation solves \eqref{GSMeqn} to give $M(\vf)$ as :
\begin{eqnarray}
	M(\vf) & = & \frac{1+2 \eta-5 \eps}{U(\vf)} \simeq - \sqrt{3\over 4V} \left[ 1 + 2 {\partial_\vf^2 V\over V}
	- {31\over 12} \left({\partial_\vf V\over V}\right)^2 \right]
\label{Msolution}
\end{eqnarray}
to first order in slow-roll.
Equations \eqref{Usolution},\eqref{Psolution}, and \eqref{Msolution} are the final results for 
the counterterms, to leading order in slow-roll parameters.

\subsection{The Power Spectrum of Slow-Roll Inflation}
\label{sec:slowroll}
The renormalized action for the gravity-scalar system is $S_{\rm tot}=S+S_{\rm ct}$, where
$S$ is the standard action \eqref{GCaction} and the counterterms are the negative of 
\eqref{sgclassicalaction}:
\begin{equation}
\label{sgcounter}
	S_{\rm ct}  = 
	- \frac{1}{8 \pi G} \, \int_{\dM} \nts \dd^3 x \sqrt{\tg} \, \left( U(\vf) + M(\vf) \vec{D} \vf \cdot \vec{D} \vf
			+ \Phi (\vf) \RR + \ldots \right)
\end{equation}
where $U(\vf)$, $M(\vf)$, and $\Phi(\vf)$ were discussed above. We now want to use this
action to analyze fluctuations around the background of slow-roll inflation:
\begin{eqnarray}
	g_{\mu\nu}(\tau) & \rightarrow & g_{\mu\nu}(\tau) + h_{\mu\nu}(\tau, \vec{x}) \\
	\vf(\tau) & \rightarrow & \vf(\tau) + \chi(\tau,\vec{x})
\end{eqnarray}
In other words, $g_{\mu\nu}$ and $\vf$ comprise the background solution, and $h_{\mu\nu}$ and $\chi$
are the fluctuations around that solution. 
Note the change in notation, where we use $\chi$ instead of $\dvf$ to refer to the fluctuation in the scalar field. 

The fluctuations of the metric $h_{\mu\nu}$ can be decomposed into scalar, vector, and tensor modes. 
We will consider only the scalar content of these fluctuations, which can be represented as:
\begin{eqnarray}
	h_{\mu\nu} & = & a^2(\tau) \, \left(\begin{array}{ccc}
				2\zeta & & D_{i} B \\
					& &	\\
				D_{i} B & & 2(\psi\,\delta_{ij} -D_{i}D_{j}E) \end{array} \right)
				\label{scalmet}
\end{eqnarray} 
When evaluating specific terms in the action we work in longitudinal gauge ($B = E = 0$). 
Thus, we are left with three scalar fields: $\zeta$, $\psi$, and $\chi$. As we evaluate the action 
we will find that these variables are related by two constraints, leaving only one physical
degree of freedom in the scalar sector.

We expand the action as a series in $\chi$ and $h_{\mu\nu}$. For convenience we work in units 
with $8 \pi G = 1$, but we will restore dimensional factors in the final answer. Indices are always 
raised and lowered with respect to the background metric $g_{\mu\nu}$. A number of results 
useful in this expansion are summarized in Appendix \ref{app:variations}. 

At zeroth order in $\chi$ and $h_{\mu\nu}$ the total action vanishes on-shell because of our definition of the counterterm action. The terms in the total action linear in $\chi$ and $h_{\mu\nu}$ are given by:
\begin{eqnarray}\label{firstorderSG}
	S_{\rm tot}^{(1)} & = & \int_{\MM_0} \bns \dd^4 x \, \sqrt{g} \, \left( \,\frac{1}{2} \, \left( T_{\mu\nu} - G_{\mu\nu} \right)\,h^{\mu\nu} 
			+ \left( \nabla^2 \vf - \partial_{\vf} V \right) \chi \, \right) \\ \nonumber
	& & - \int_{\dM_0} \bns \dd \vec{x} \, \sqrt{\tg} \, \left( \left( P_{\vf} - \pi_{\phi} \right) \, \chi + \left( P_{ij} - \pi_{ij} \right) 
			\, h^{ij} \right) 
\end{eqnarray}
In the first term $T_{\mu\nu}$ is the energy-momentum tensor and $G_{\mu\nu}$ is the Einstein tensor. The condition for this term to vanish on-shell is Einstein's equation:
\begin{eqnarray}
	G_{\mu\nu} & = & 8 \pi G \, T_{\mu\nu}
	\label{einstein}
\end{eqnarray}
The second term represents the equation of motion for $\vf$ which also vanishes on-shell. 
This leaves the boundary terms, which we have written using a compact notation:
\begin{eqnarray}
	P_{\vf} & = & \partial_{\vf} U - \partial_{\vf} M \, \vec{D} \vf \cdot \vec{D} \vf - 2 M \vec{D}^2 \vf + \RR \, \partial_{\vf} \Phi \\
	P_{ij} & = & \frac{1}{2} \, g_{ij} \left( U + M \vec{D} \vf \cdot \vec{D} \vf \right) - M D_{i} \vf D_{j} \vf 
			-  \, \Phi \, \GG_{ij} + D_{i} D_{j} \Phi - g_{ij} \vec{D}^2 \Phi
\end{eqnarray}
These expressions are simply equations \eqref{HJmetricmomentum} and \eqref{HJscalarmomentum} renamed, and are equal to the canonical momenta $\pi_{ij}$ and $\pi_{\phi}$ evaluated at the boundary $\dM_0$. As a result the boundary terms cancel on-shell, and the first order term in the action vanishes.

The first non-zero contributions to the action appear at quadratic order. They are computed by 
varying the first order action \eqref{firstorderSG} first, and only then 
imposing the background equation of motion. The bulk terms are given by:
\begin{eqnarray}
	S_{\rm bulk}^{\,(2)} & = &  \int_{\MM_0} \bns \dd^4 x \, \sqrt{g} \, 
	\left( \,\frac{1}{2} \, H_{\mu\nu} \,h^{\mu\nu} 
	+ \delta \left( \nabla^2 \vf - \partial_{\vf} V \right) \, \chi \right)
	\label{bulkt}
\end{eqnarray}
where $H_{\mu\nu}$ is:
\begin{eqnarray}
	H_{\mu\nu} & = & \delta T_{\mu\nu} - \delta G_{\mu\nu}
\end{eqnarray}
The explicit expressions for $H_{\mu\nu}$
are given in appendix \ref{app:variations}.

The bulk equations of motion of the full system are Einstein's equation \eqref{einstein}
and the Klein-Gordon equation. They are satisfied by the background and by the
total configuration separately, and therefore also by the fluctuations. The bulk equations
of motion for the fluctuations are therefore $H_{\mu\nu}=0$. We will not analyze the
corresponding equation for the scalar field $\delta \left( \nabla^2 \vf - \partial_{\vf} V \right)=0$
here since it is redundant. 

In the longitudinal gauge $E=B=0$. The corresponding equations of motion
$H_{\tau i}=0$ and $H_{ij}-{1\over 3}\delta_{ij}H_k^k=0$ are therefore constraints. 
These constraints will each remove one degree of freedom, leaving just one physical scalar.
The E-constraint $H_{ij}-{1\over 3}\delta_{ij}H_k^k=0$ implies:
\begin{eqnarray}
\label{zetapsiconstraint}
	\zeta & = & \psi
\end{eqnarray}
We will enforce this constraint and keep only $\zeta$ in the rest of the calculation. 
The B-constraint $H_{\tau i} = 0$ expresses $\chi$ as:
\begin{eqnarray}\label{zetachiconstraint}
	\zeta\,' + \HH \,  \zeta + \frac{1}{2} \, \vf\,' \, \chi & = & 0
\end{eqnarray}
We will take $\zeta$ to represent the single scalar degree of freedom in the problem and eventually 
express the total action as a function of $\zeta$ only. However, the rest of the calculation is 
simplified by keeping both $\chi$ and $\zeta$ with the understanding
that \eqref{zetachiconstraint} can be imposed when needed.

The remaining bulk equations $H_{\tau\tau}=0$ and $H_k^k=0$ constitute a coupled set of equations 
of motion for $\zeta$ and $\chi$. Using the constraint \eqref{zetachiconstraint} they can be
disentagled to give just one equation of motion for $\zeta$:
\begin{eqnarray}
	\zeta\,'' + 2 \left( \HH - \frac{\vf\,''}{\vf\,'} \right) \, \zeta\,' - \vec{\partial}^2 \zeta + 
		2 \left( \HH\,' - \HH\, \frac{\vf\,''}{\vf\,'}\right)\, \zeta & = & 0
\end{eqnarray}
This equation is valid for arbitrary FRW spacetimes. For inflationary space-times satisfying 
the slow-roll conditions  it becomes:
\begin{eqnarray}
	\zeta\,'' - 2 \frac{\left( \eta-\eps\right)}{\tau} \, \zeta\,' - \vec{\partial}^2 \zeta + 2 \frac{\left( \eta-2\eps\right)}{\tau^2} \, \zeta & = & 0
	\label{sloweom}
\end{eqnarray}

So far we have considered the second variation of the bulk terms only. These give the
equations of motion, but do not contribute to the H-J functional since \eqref{bulkt} vanishes 
on-shell, by definition. The second variation of the boundary terms, found by varying 
\eqref{firstorderSG}, are given by:
\begin{eqnarray}\label{GSquadbndy}
	S_{\rm bndy}^{\,(2)} & = &  \, \int_{\dM_0} \bns  \dd^3 x \sqrt{\tg} \, \left( \left( \delta P_{\vf} - \delta \pi_{\vf} \right) \, \chi + 
			\left( \delta P_{ij} - \delta \pi_{ij} \right) h^{ij} \right)
\end{eqnarray}
We have computed these variations for arbitrary backgrounds and present the result in 
Appendix \eqref{momenvar}. For a background which is a flat FRW space they are:
\begin{eqnarray}
\label{varPvf}
	\delta P_{\vf}  & = & \partial_{\vf}^{\,2} U \, \chi - 2 M \vec{D}^2 \chi - 4 \partial_{\vf} \Phi \, \vec{D}^2\zeta  \\
	\delta \pi_{\vf} & = & -\frac{\vf\,'}{a}  \, \zeta - \frac{1}{a} \,  \,\chi\,' \\
	\delta P_{ij} & = & \frac{1}{2} \, g_{ij} \partial_{\vf} U \, \chi + \partial_{\vf} \Phi \, \left( D_i D_j \chi - g_{ij} \vec{D}^2 \chi \right)
		+ 3 U \zeta + \frac{1}{2} \, \Phi \left( D_i D_j \zeta - g_{ij} \vec{D}^2 \zeta \right) \\
	\delta \pi_{ij} & = & \frac{1}{4} \, h_{j}^{\,k} K_{ik} - \frac{1}{2} \, h_{ij} K
		+ \frac{1}{4} \, g_{ij} \, h^{\lambda \rho} K_{\lambda\rho} \label{varpij}
		 \\ \nonumber
		& & \, \,+ \frac{1}{4} \, n_{\lambda} \left( \nabla_{j} h^{\lambda}_{i} - \nabla^{\lambda} h_{ij}\right)
			- \frac{1}{4} \, g_{ij} n_{\lambda} \left( \nabla_{\rho} h^{\lambda \rho} - \nabla^{\lambda} h^{\rho}_{\,\rho}\right) 
\end{eqnarray}
Using these expressions to evaluate the action \eqref{GSquadbndy} gives $S_{\rm tot}$ to 
quadratic order in $\zeta$ and $\chi$:
\begin{eqnarray} \nonumber
	S_{\rm tot} & = & \frac{1}{2}\,\int_{\dM_0} \bns \dd^3 x \sqrt{\tg} \left( -\frac{6}{a} \, \zeta \zeta \,' - 6 \left( U + 3 \frac{\HH}{a} \right) \zeta^2 
		+ 4\,\Phi \, \zeta \vec{D}^2 \zeta + \frac{1}{a} \, \chi \chi\,'  + 2 M \, \chi \vec{D}^2 \chi \right. \\ 
		& & \hspace{20pt} \left.  - \,\partial_{\vf}^{\,2} U \, \chi^2  
			+ \left( \frac{\vf\,'}{a} - 3 \,\partial_{\vf} U \right) \zeta \chi +4 \,\partial_{\vf} \Phi \, 
			\left( \chi \vec{D}^2   \zeta + \zeta \vec{D}^2 \chi  \right) \right)
\end{eqnarray}
Using the constraint \eqref{zetachiconstraint}, the equation of motion for $\zeta$, and the expression
\eqref{Umaster} for $U(\vf)$ this simplifies to:
\begin{eqnarray}
	S_{\rm tot} & = & \frac{1}{2}\,\int_{\dM_0} \dd^3 x \sqrt{\tg} \, \left( - \frac{2 a }{ \, \vf\,'} \, \chi 
	\vec{D}^{\,2} \zeta + 2 M \, \chi \vec{D}^{\,2} \chi + 
		4 \Phi \, \zeta \vec{D}^{\,2} \zeta + 8 \, \partial_{\vf} \Phi \, \chi \vec{D}^2 \zeta\right)
\end{eqnarray}
This equation is valid for general FRW backgrounds. 

We now specialize to slow-roll inflation, 
use \eqref{Psolution} and \eqref{Msolution} for $\Phi(\vf)$ and $M(\vf)$, and eliminate $\chi$ 
using the constraint \eqref{zetachiconstraint}. Writing the result in Fourier space, we obtain:
\begin{eqnarray}
	S_{\rm tot} & = & \int \dd^3 k \,\dd^3 p \, \delta^{(3)}(\vec{k}+\vec{p}) \, \frac{2\vec{k}^{\,2}}{(\vf\,')^2} \, 
		\left( \frac{a^2}{\HH} ( 1+2 \eta - 5 \eps)\zeta_{\vec{k}}\,' \zeta_{\vec{p}}\,'  + \right.\\ \nonumber & & \,\,\,\,\,\,\,\,\left. \phantom{\frac{a^2}{\HH}} 
		a^2 \left( 1+ 4( \eta - 2 \eps )\right) \zeta_{\vec{k}}\,' \, 
			\zeta_{\vec{p}} + 2\,a^2 \, \HH \, (\eta-2 \eps )
			\,\zeta_{\vec{k}} \, \zeta_{\vec{p}}  
			\right)
			\label{finacts}
\end{eqnarray}
This is the H-J functional in terms of the physical field $\zeta$ and its time derivative. 
We must now use the bulk equations of motion to eliminate the time derivative as we did
previously, {\it e.g.} in rewriting \eqref{onsact}. The equations of motion \eqref{sloweom}
and the regularity condition at $\tau\to -\infty$ gives the solution:
\begin{equation}
\label{zetasol}
\zeta = |\tau|^{{1\over 2}+\eta -\eps}\left( \phantom{\frac{}{}}  
J_{-\nu}(|k\tau|) - e^{i\pi\nu}J_{\nu}(|k\tau|) \right)
\end{equation}
where $\nu={1\over 2}- \eta + 3\eps$. Note that this expression for $\nu$, and the exponent of
$\tau$ in the prefactor of \eqref{zetasol}, are approximately ${1\over 2}$ in the present case,
where gravity is taken into account. In the corresponding expression \eqref{posfreqsoln}
for a fixed background both were approximately ${3\over 2}$. The overall normalization of
\eqref{zetasol} is not relevant for our purposes.

The explicit solution for $\zeta$ allows us to rewrite the terms involving time 
derivatives, following the steps leading to \eqref{finalF}. Inserting the result 
in \eqref{finacts} we find the
quadratic action for the fluctuations:
\begin{eqnarray} \label{GSfinalaction}
	S_{\rm tot} & = & -2 \int \dd^3 k \, \left(\frac{\HH}{\,\vf\,'}\right)^2 \frac{ik^3}{\,H^2} 
	(k \tau_0)^{6 \eps - 2 \eta}   \, \zeta_{\vec{k}} \, \zeta_{-\vec{k}}
\end{eqnarray}
There are no power-law divergences in this result. They appear in intermediate steps
but they cancel between the bare terms $\delta \pi_{\vf}, \delta \pi_{ij}$ and
the counterterms $\delta P_{\vf}, \delta P_{ij}$. These cancellations provide a detailed
check of our procedure and out computations.

The semiclassical wave function: 
\begin{eqnarray}
	\Psi [\zeta] & = & \exp \left( i S_{\rm tot} [\zeta]\right)
\end{eqnarray}
gives the two-point correlation function:
\begin{eqnarray}
\label{zetacorr}
	\langle \zeta_{\vec{k}} \,\zeta_{-\vec{k}} \rangle & = & 
	\int \DD \zeta \,\, \zeta_{\vec{k}} \, \zeta_{-\vec{k}} \, |\Psi[\zeta]|^2=
	\eps \, \frac{H^2}{4 k^3} \, (k \tau_0)^{2\eta-6\eps}
\end{eqnarray}
The corresponding power spectrum for $\zeta$ is then given by:
\begin{eqnarray}
	P_{\zeta}(\vec{k}) & = & \frac{\eps}{2} \, \left( \frac{H}{2\pi} \,\right)^2 \, 
	\left( k \tau_0 \right)^{2\eta -6\eps}
\end{eqnarray}
The overall factor of $\eps$ in the correlator comes from the factor of $(\HH / \vf')^2$ in 
\eqref{GSfinalaction}. This factor can be understood by noting that the limit $\eps\to 0$ 
corresponds to $\pvf V \to 0$, a constant potential. In this limit the spacetime is 
(asymptotically) de Sitter space, and the scalar flucutations in the metric are pure gauge. 
The factor of $\eps$ ensures that the power spectrum for $\zeta$ vanishes in the $\eps \to 0$
limit as expected.

The standard result presented in the inflation literature is the power spectrum for 
perturbations in the spatial curvature $\RR$. This quantity is useful because it 
stays constant while on superhorizon scales and later transfers rather directly
into the observable perturbations in matter and radiation. 
The comoving curvature perturbations are given by:
\begin{eqnarray}
	\RR & = &- \frac{\HH}{\,\vf\,'} \, \chi = \frac{1}{\eps} \,\zeta
	\label{Rdef}
\end{eqnarray}
for small $\eps$. The slow roll parameter in the denominator translates to a
crucial enhancement of the density perturbations. Combining \eqref{Rdef} with
the power spectrum for $\zeta$ and \eqref{GSeps} for the slow roll parameter we 
find:
\begin{eqnarray}
	P_{\RR} (\vec{k}) & = & \left( {H\over {\dot{\vf}}}\right)^2~ \left( \frac{H}{2\pi} \, \right)^2 \, \left( k \tau_0 \right)^{2\eta - 6 \eps}
	\label{finR}
\end{eqnarray}
The scaling is usually characterized by the spectral index $n_s$ which is $1$ for
scale invariant perturbations. Our result is
\begin{equation}
n_s - 1 = 2\eta-6\eps  
\end{equation}
This agrees with more conventional computations (see {\it e.g.} ~\cite{Liddle:cg,Peacock:ye,Peebles:xt}).

We have written our results for the renormalized action, and for the correlators of $\zeta$
and ${\cal R}$, with the explicit cut-off retained. As we discussed in section \ref{logdivs},
the cutoff can in fact be removed, if one should wish to do so. In the present case
the FRW equations allow us to estimate the scalings 
$H\sim\tau_0^\eps$, $\eps\sim\tau_0^{2\eta-4\eps}$ and \eqref{zetasol} gives $\zeta\sim\tau_0^{2\eta-4\eps}$. These show that \eqref{GSfinalaction} 
and \eqref{finR} are independent of $\tau_0$ and that \eqref{zetacorr} scales like
two $\zeta$'s.  This serves as a check on our computations. However, again, it is 
physically more appropriate to simply take the physical cutoff $\tau_0=1/aH$. 

\section{The Power Spectrum from the Renormalization Group}
\label{sec:rgcft}
In this section we discuss some aspects of the holographic renormalization group, {\it i.e.}
we explore the relation between inflationary physics and a conjectured three-dimensional 
quantum field theory near its renormalization group fixed point. Alternatively, the considerations 
can be interpreted conservatively, as comments on abstract properties of the IR-divergences.

\subsection{The RG Equation}
There have been many attempts to formulate a holographic duality between a three dimensional
quantum field theory and gravitational physics in de Sitter space, or inflationary spacetimes, including \cite{Strominger:2001pn,Strominger:2001gp,Witten:2001kn,Larsen:2002et,Maldacena:2002vr,Freedman:1999gp,Balasubramanian:zh,Balasubramanian:2001nb}. 
Interpretations vary in part because, unlike AdS/CFT, there is no candidate microscopic description. For the purposes of this section we will simply assume that
such a duality exists and discuss a few of its properties. 

The H-J functional depends on the fields at some late time $\tau_0$, which we have 
interpreted as the infrared cut-off. Holography reinterprets the H-J functional as the
effective action of a three dimensional quantum field theory with the bulk field now playing the
role of couplings. The unperturbed theory is thought to be conformal so $\vf$ deforms 
the theory as: 
%~\footnote{For notational simplicity we consider a field in a FRW background
%but the considerations apply to the full gravity scalar system discussed in 
%section \ref{sec:gravscalar} as well. }: 
\begin{eqnarray}\label{deformation}
	\Lag_{CFT} & \rightarrow & \Lag_{CFT} + \vf \, \OO
\end{eqnarray}
The time $\tau_0$ becomes the scale in the boundary theory via $\mu \propto a(\tau_0)$, 
which famously relates UV and IR physics. The $\beta$-function associated with the 
coupling in the deformation \eqref{deformation} is:
\begin{eqnarray} \label{betadefinition}
	\beta & = & \frac{\partial \,\vf}{\partial \log{a}}
\end{eqnarray}
Since the background field $\vf$, which we are now interpreting as a coupling, 
may have spatial dependence we are really considering a $\beta$-functional here. 

Our goal here is to derive a simple differential equation satisfied by the renormalized action
$S_{\rm tot}$:
\begin{eqnarray}\label{originalaction}
	S_{\rm tot} & = & S - S_{\rm ct}
\end{eqnarray}
The ingredient we wish to exploit is, once again, the H-J equation. The 
H-J equation \eqref{scalargravityHJ} is satisfied by the full H-J functional $S$
but, by construction, it is also satisfied by the counter-term Lagrangian 
$S_{\rm ct}$, at least as an expansion valid for small values of the cut-off $\tau_0$. 
To exploit this, we decompose the canonical momenta of $S$ into two terms, one due 
to $S_{\rm tot}$ and one due to $S_{\rm ct}$:
\begin{eqnarray}
	\pi_{\vf} & = & {\rm P}_{\vf} + \mathcal{P}_{\vf} 
\end{eqnarray}
where:
\begin{eqnarray}
	{\rm P}_{\vf} \, = \, \frac{1}{\sqrt{\tg}} \, \frac{\delta S_{\rm ct}}{\delta \vf} & & 
		\mathcal{P}_{\vf} \, = \, \frac{1}{\sqrt{\tg}} \, \frac{\delta S_{\rm tot}}{\delta \vf}
\end{eqnarray}
We make a similar decomposition for $\pi^{ij}$. Substituting these expressions in the H-J eqn \eqref{scalargravityHJ} we find:
\begin{eqnarray}
	4 \left( {\rm P}^{ij} \, \mathcal{P}_{ij} - {1\over 2} \, {\rm P}^{i}_{\,i} \,\mathcal{P}^{j}_{\,j} \right) + {\rm P}_{\vf} \,\mathcal{P}_{\vf} 
	+ 2 \left( {\cal P}^{ij} \, {\cal P}_{ij} - {1\over 2} \, {\cal P}^{i}_{\,i} \,{\cal P}^{j}_{\,j} \right) + \frac{1}{2} {\cal P}_{\vf}^{\,2} & = & 0
\end{eqnarray}
We are interested in the leading terms in $S_{\rm tot}$ and so assume that the terms quadratic in 
${\cal P}_{\vf}$ and ${\cal P}_{ij}$ are negligible compared to the linear terms. This leaves the 
equation:
\begin{eqnarray}\label{firstRGeqn}
	4 \left( {\rm P}^{ij}  - {1\over 2} \, {\rm P}^{k}_{\,k} \, g^{ij}  \right) \frac{\delta S_{\rm tot}}{\delta g^{ij}} + {\rm P}_{\vf} \,
		\frac{\delta S_{\rm tot}}{\delta \vf}  & = & 0
\end{eqnarray}
This constitutes a linear differential equation for $S_{\rm tot}$ whose coefficients, 
given by functional derivatives of $S_{\rm ct}$, for general backgrounds depend on the 
functions $U$, $\Phi$, and $M$. 

If the background is a spatially flat FRW cosmology the equations \eqref{HJscalarmomentum} and \eqref{HJmetricmomentum} for the functional derivatives of $S_{\rm ct}$ specify the coefficients in  \eqref{firstRGeqn} which becomes:
\begin{eqnarray}
\label{nextrgeqn}
	g_{ij} \, \frac{\delta S_{\rm tot}}{\delta g_{ij}}  - \frac{\pvf U}{U} \, \frac{\delta S_{\rm tot}}{\delta \vf}& = & 0
\end{eqnarray}
The variations with respect to the metric amount to overall changes in the
scale factor $a(\tau)$:
\begin{eqnarray}
	g_{ij} \, \frac{\,\delta\,}{\delta g_{ij}} & = & \frac{1}{2} \, a\, \frac{\delta}{\delta a}
\end{eqnarray}
Additionally, the definition of the $\beta$-function \eqref{betadefinition} gives:
\begin{eqnarray}
	\beta & = & \frac{1}{H}\pi_\vf = \frac{1}{H}\partial_\vf U = -2 \, \frac{\pvf U}{U}
\end{eqnarray}
using \eqref{HJmomentum} and \eqref{Umaster}; so the coefficient of the second 
term is the $\beta$-function. Substituting in \eqref{nextrgeqn} we find:
\begin{eqnarray}\label{RGeqn}
	\left( \frac{\partial}{\partial \log{a}} + \beta \, \frac{\delta}{\delta \vf} \right) \, S_{\rm tot} & = & 0
\end{eqnarray}
This we can view as an RG equation for the renormalized action. In section \ref{sec:gravscalar} we calculated the action $S_{\rm tot}$. The fact that our expression \eqref{GSfinalaction} for the action satisfies \eqref{RGeqn} is a non-trivial consistency check of our approach.

\subsection{The Callan-Symanzik Equation}

In the introduction we alluded to the fact that the holographic interpretation of our approach is conceptually different than the standard
approach to holographic RG flows in AdS/CFT. The idea is that we would think of the evolution of an inflationary spacetime as a flow in
the space of three-dimensional theories. We can use the RG equation \eqref{RGeqn} to make a more precise version of this statement by
relating measuarable quantities in inflation to the quantities characterizing the RG flow in the dual theory.

Following \cite{Maldacena:2002vr}, we treat the exponential of the on-shell action $S_{\rm tot}$ as the generating function of correlators in the dual theory:
\begin{eqnarray}\label{genfunction}
	Z[\vf] & \sim & \exp{(i S_{\rm tot})}
\end{eqnarray}
The boundary data for the bulk field is interpreted as a source for operators that we schematically denote $\OO(\vec{x})$, so that we can obtain correlators of the $\OO(\vec{x})$ by taking functional derivatives:
\begin{eqnarray} \label{2pointfunction}
	\langle \OO(\vec{x}) \OO(\vec{y}) \rangle & = & \frac{\delta^2 S_{\rm tot}}{\delta \vf(\vec{x}) \delta \vf(\vec{y})}
\end{eqnarray}
Applying functional derivatives to the RG equation \eqref{RGeqn} and integrating the result over the three-dimensional space, we arrive
at a Callan-Symanzik equation satisfied by the correlators
of operators at distinct points:
\begin{eqnarray}\label{CSeqn}
	\left(  a \frac{\partial}{\partial a} + \beta(\vf) \, \frac{\partial}{\partial \vf} + n \, \gamma(\vf) \right) \, 
		\langle \OO(\vec{x}_1) \ldots \OO(\vec{x}_n) \rangle & = & 0
\end{eqnarray}
The third term in the equation contains the anomalous dimension $\gamma$, which is defined~\footnote{ Note that this is not the standard definition of the
anomalous dimension of $\OO$, which is usually defined as $\gamma = \frac{\partial \log{\vf}}{\partial \log{a}} = \beta / \vf$.} as:
\begin{eqnarray}
	\gamma = \pvf \beta
\end{eqnarray}
The construction of the Callan-Symanzik equation is technically similar to examples in AdS/CFT \cite{deBoer:1999xf,Bianchi:2001kw}. 

We can now use the Callan-Symanzik equation \eqref{CSeqn} to derive an expression for the spectral index of inflation in terms of the functions
describing the RG flow. The important point is that the generating function defined in equation \eqref{genfunction} is essentially the semi-classical wave function of the Universe. If we take the action $S_{\rm tot}$ to be quadratic then the two-point correlator of $\OO$ is related to the two-point function of the bulk mode $\zeta_{\vec{k}}$ by:
\begin{eqnarray}\label{relnbetweencorrelators}
	\langle \OO_{\vec{k}} \, \OO_{-\vec{k}} \rangle & \sim & \frac{\eps}{\langle \zeta_{\vec{k}} \,\zeta_{-\vec{k}}\rangle}
\end{eqnarray}
The factor of $\eps$ in the numerator comes from equation \eqref{Rdef}, which tells us that $\zeta \sim \sqrt{\eps} \, \chi$. The spectral index is defined as:
\begin{eqnarray}
	n_s - 1 & = & k \, \frac{d}{d k} \log{\left(k^3 \langle \zeta_{\vec{k}} \,\zeta_{-\vec{k}} \rangle\right)}
\end{eqnarray}
Using equation \eqref{relnbetweencorrelators} and the expression \eqref{zetacorr} 
for the two-point function of $\zeta_{\vec{k}}$ we can
rewrite the definition of the spectral index as:
\begin{eqnarray}
	n_s - 1 & = & a \, \frac{d}{d a} \log{ \langle \OO_{\vec{k}} \,\OO_{-\vec{k}} \rangle}
\end{eqnarray}
This gives us the first term in the C-S equation. We can evaluate the second term in the same way, which gives:
\begin{eqnarray}
	\beta \, \frac{\partial}{\partial \vf} \,  \langle \OO_{\vec{k}} \,\OO_{-\vec{k}} \rangle & = &  \beta^{\,2} \,  \langle \OO_{\vec{k}} \,\OO_{-\vec{k}} \rangle
\end{eqnarray}
Using these results, the Callan-Symanzik equation reduces to an expression for the spectral index in terms of $\beta$ and $\gamma$:
\begin{eqnarray}\label{spectralindexeqn}
	n_s  & = & 1 - \beta^{\,2} - 2 \gamma 
\end{eqnarray}
Using the slow-roll result for $U(\vf)$, equation \eqref{Usolution}, and the definition for $\beta$, one can verify that:
\begin{eqnarray}
	\beta^{\,2} & = & 2 \eps \\
	\gamma & = & 2 \eps - \eta
\end{eqnarray}
Evaluating equation \eqref{spectralindexeqn} using these expressions reproduces the standard slow-roll result:
\begin{eqnarray}
	n_s & = & 1 + 2 \eta - 6 \eps
\end{eqnarray}
It is very interesting that this expression, possibly the one explaining the observable
cosmic density perturbations, can be derived fairly straightforwardly from holographic
ideas. We anticipate that more general expressions, valid to all orders in slow-roll 
(see {\it e.g.} \cite{Copeland:1993jj,Kinney:2002qn}) allow a similar holographic 
interpretation. Of course such expressions also allow the more conservative
interpretations as the renormalization groups controlling the infrared behavior
of gravity.

Let us conclude with a philosophical comment:
the framework advocated here represents a step towards formulating inflation in terms that 
emphasize symmetries, particularly scaling symmetries, rather than the customary focus on 
specific inflationary potentials. We interpret inflation as a phase of quantum gravity with a 
special symmetry, the scaling invariance of pure de~Sitter space. This symmetry is broken of 
course, since inflationary spacetimes are only approximately de~Sitter, but the broken 
symmetry can be treated in perturbation theory around the scaling solution. Thus inflation 
is interpreted as broken scale invariance. 

%Suppose inflation emerged from an asymptotically de~Sitter space in the
%infinite past, through some perturbation. In such a cosmology, de~Sitter
%space can be interpreted holographically as the dual conformal field
%theory, and the perturbation starting the universe is the addition of
%an irrelevant operator to the Lagrangean. 

%Since our results coincide with the standard inflationary predictions we obviously
%agree with observations as well. However, the interpretation of inflation as broken
%scale invariance implies that de~Sitter space is an {\it attractive} fixed point and,
%as noted in~\cite{Larsen:2002et}, this imposes the constraint that the spectrum is 
%red in the deep infra-red, {\it i.e.} the spectral index $n_s<1$ for small multipole $\l$. 
%The recent WMAP observations \cite{Bennett:2003bz} marginally contradicts this. In fact they suggest a %shift from 
%red to blue spectrum close to the infrared. If true, this would be interpreted in inflation as a 
%complicated inflationary potential. In our approach this behavior corresponds to 
%an attractive fixed point perturbed by a relevant operator with exceptionally small 
%coefficient. This is not an aesthetically pleasing model, but the example indicates that an
%approach based on scaling violations could be a useful guide to interpreting the
%apparently bizarre world.

\section*{Acknowledgements}
We thank J. de Boer, V. Husain, N. Kaloper, W. Kinney, and D. Minic for discussions
and DoE for financial support. 

\pagebreak

\appendix
\section{Some General Formulae} \label{app:Conventions}

We consider four dimensional spacetimes $\MM$ with a spacelike boundary $\dM$ and metric $g_{\mu\nu}$. The spacetime is foliated by a family of spacelike hypersurfaces orthogonal to a timelike unit normal vector $n^{\mu}$. Specifically, the spacelike hypersurfaces are taken to be surfaces of constant time, although we have not specified any particular system of coordinates. In that case one can think of $n^{\mu}$ as the four velocity of an observer moving orthogonally to the constant time hypersurfaces. There is also a four-acceleration given by:
\begin{eqnarray}
	a^{\mu} & = & n^{\nu} \nabla_{\nu} \,n^{\mu}
\end{eqnarray}
Four dimensional tensors are projected onto the spatial hypersurfaces using the projection tensor:
\begin{eqnarray}
	P_{\mu}^{\,\nu} & = & \delta_{\mu}^{\,\nu} + n_{\mu} n^{\nu}
\end{eqnarray}
The projection of an arbitrary tensor $T^{\mu \ldots}_{\,\,\nu \ldots}$ is given by:
\begin{eqnarray}
	P \, T^{\,\mu \ldots}_{\,\,\nu \ldots} & = & P^{\mu}_{\,\lambda} \, P^{\rho}_{\,\nu} \, \ldots \, T^{\,\lambda \ldots}_{\,\,\rho \ldots}
\end{eqnarray}
The resulting tensor is completely spacelike and orthogonal to $n^{\mu}$ in all its indices. The induced metric on one of the constant time hypersurfaces is denoted $\tg_{\mu\nu}$ and is given by the projection of the metric $g_{\mu\nu}$:
\begin{eqnarray}\label{inducedmetric}
	\tg_{\mu\nu} & = & g_{\mu\nu} + n_{\mu} n_{\nu}
\end{eqnarray}
The induced metric inherits a covariant derivative $D_{\mu}$ from the four dimensional covariant derivative $\nabla_{\mu}$, defined by the relation:
\begin{eqnarray}
	D_{\mu} T^{\ldots}_{\ldots} & = & P \,\nabla_{\mu} T^{\ldots}_{\ldots}
\end{eqnarray}
Similarly, there is an an intrinsic Riemann tensor associated with the induced metric that is defined by the commutator of covariant derivatives on an arbitrary spacelike vector $A_{\mu}$:
\begin{eqnarray}
	[D_{\mu}\,, D_{\nu}] A_{\lambda} & = & \RR^{\rho}_{\,\,\lambda \,\nu \mu} \, A_{\rho}
\end{eqnarray}
The corresponding Ricci tensor and scalar are:
\begin{eqnarray}
	\RR_{\mu\nu} & = & \delta^{\lambda}_{\,\rho} \, \RR^{\rho}_{\,\,\mu  \lambda \nu} \\
	\RR & = & g^{\mu\nu} \, \RR_{\mu\nu}
\end{eqnarray}
In addition to the intrinsic curvature tensors defined above, there is an extrinisic curvature that characterizes how each spacelike hypersurface is embedded in $\MM$. It is given by:
\begin{eqnarray}
	K_{\mu\nu} & = & - P \, \nabla_{(\mu} n_{\nu)}  \\
			    & = & - \frac{1}{2} \left( \nabla_{\mu} n_{\nu} + \nabla_{\nu} n_{\mu} + n_{\mu} a_{\nu} +  n_{\nu} a_{\mu}\right) 
\end{eqnarray}

Applying the projection tensor to various contractions of the four dimensional Riemann tensor allows us to derive a number of useful identities.
Collectively referred to as the \emph{Gauss-Codazzi} equations, they allow us to rewrite projections of four dimensional curvature tensors in terms of the intrinsic and extrinsic curvatures defined above. The main equations are:
\begin{eqnarray}
	P \, R_{\mu\nu\lambda\rho} & = & \RR_{\mu\nu\lambda\rho} + K_{\rho\nu} \, K_{\lambda \mu} - K_{\rho\mu} \, K_{\lambda \nu} \\
	P \, \left( n^{\mu}  R_{\mu\nu\lambda\rho} \right) & = & D_{\rho} K_{\lambda \nu} - D_{\lambda} K_{\rho\nu}
\end{eqnarray}From these equations we can obtain the following useful identities:
\begin{eqnarray} \label{GCone}
	n^{\mu} n^{\nu} R_{\mu\nu} & = & K^2 - K^{\mu\nu} K_{\mu\nu} + \nabla_{\mu} \left( a^{\mu} + n^{\mu} K \right)
\end{eqnarray}
\begin{eqnarray} \label{GQtwo}
	P \, R_{\mu \nu} n^{\mu} & = & D_{\mu} K_{\,\,\nu}^{\mu} -  D_{\nu} K 
\end{eqnarray}
\begin{eqnarray} \label{GCthree}
	R = \RR + K^{\mu\nu} K_{\mu\nu} - K^2 - 2 \,\nabla_{\mu} \left(  a^{\mu} + n^{\mu} K\right)
\end{eqnarray}

\subsection{Explicit Computations}
The calculations in section \ref{sec:gravscalar} require us to evaluate many of these tensors for a metric of the form:
\begin{eqnarray}\label{metric}
	d s^2 & = & a(\tau)^2 \, \left( -d \tau^2 + \gamma_{ij} (\vec{x}) d x^i d x^j \right)
\end{eqnarray}
Given the explicit 3+1 split of the  metric in these coordinates we use greek letters $\mu,\nu,\ldots$ for four dimensional (spacetime) indices, and roman letters $i,j,\ldots$ for spatial indices. The time $\tau$ is the conformal time, and we use primes to denote derivatives with respect to it. We will also use the abbreviation $\HH = \frac{\,a\,'}{a}$. 

The timelike normal is given by $n_{\mu} = a(\tau) \delta _{\mu\tau}$~\footnote{This is a non-standard choice of $n^{\mu}$, as it corresponds to a \emph{past-directed} timelike unit normal vector. The consequences consist entirely of an occasional difference in sign compared to the standard convention. }, and the metric
induced on constant $\tau$ hypersurfaces is just the spatial part of the metric:
\begin{eqnarray}
	\tg_{\mu\nu} & = & g_{ij} \, \delta_{\mu i} \,\delta_{\nu j}
\end{eqnarray}
The non-zero components of the Ricci tensor are:
\begin{eqnarray}
	R_{\tau\tau} & = & -3 \, \HH \,' \\
	R_{i j} & = & \RR_{ij} + 2 \left(\,\frac{\HH}{a}\,\right)^2 \, g_{ij} + \frac{\,\HH\,'}{a^2} \, g_{ij} 
\end{eqnarray}
And the Ricci scalar is:
\begin{eqnarray}
	R & = & \RR + 6 \left(\,\frac{\HH}{a}\,\right)^2 + 6\,\frac{\,\HH\,'}{a^2}
\end{eqnarray}
Note that the spatial curvature of the spatial slice is defined with respect to the full metric
so that
\begin{eqnarray}
	\RR & = & \frac{1}{\,a^2} \, R(\gamma_{ij})
\end{eqnarray}
The extrinsic curvature is given by:
\begin{eqnarray}
	K_{ij} & = & \frac{\HH}{a} \, g_{ij}
\end{eqnarray}

\section{Variations and Functional Derivatives}\label{app:variations}
In section \eqref{sec:gravscalar} we consider small fluctuations around  background fields 
$g_{\mu\nu}$ and $\vf$:
\begin{eqnarray}
	g_{\mu\nu} & \rightarrow & g_{\mu\nu} + h_{\mu\nu} \\ \nonumber
	\vf  & \rightarrow & \vf + \chi
\end{eqnarray}
Schematically, we expand functions of the fields as a Taylor-series:
\begin{eqnarray}
	\FF(\vf+\chi) & = & \FF(\vf) + \delta \FF(\vf) \, \chi +  \frac{1}{2} \, \delta^2 \FF(\vf) \, \chi^2 + \ldots
\end{eqnarray}
In this expression $\delta$ is an operator that `linearizes' the function it acts on. For example:
\begin{eqnarray}
	\delta g_{\mu\nu} & = & h_{\mu\nu} \\ \nonumber
	\delta \vf & = & \chi \\ \nonumber
	\delta \vf^{\,2} & = & 2 \,\vf \,\chi
\end{eqnarray}
Applying $\delta$ to $g_{\mu\nu}$ or $\vf$ more than once gives zero:
\begin{eqnarray}
	\delta^2 g_{\mu\nu} & = & 0 \\ \nonumber
	\delta^2 \vf & = & 0
\end{eqnarray}

\subsection{Variations of Tensors}

Evaluating variations and functional derivatives with respect to $\vf$ is straightforward. On the other hand, expanding tensors which are functions of the metric can be non-trivial. The following expressions are 
useful:
\begin{eqnarray}
	\delta  g^{\mu\nu} & = & - h^{\mu\nu} \\
	\delta \sqrt{g} & = & \frac{1}{2} \, \sqrt{g} \, h^{\mu\nu} g_{\mu\nu} \\
	\delta \Gamma^{\lambda}_{\mu\nu} & = & \frac{1}{2} \, \left( \nabla_{\mu} h^{\lambda}_{\,\nu} + 
			\nabla_{\nu} h^{\lambda}_{\,\mu} - \nabla^{\lambda} h_{\mu\nu} \right) \\ \label{deltaRmunu}
	\delta R_{\mu\nu} & = & \frac{1}{2} \, \left( \nabla_{\lambda} \nabla_{\mu} h^{\lambda}_{\,\nu} + 
			\nabla_{\lambda} \nabla_{\nu} h^{\lambda}_{\,\mu} - \nabla_{\mu} \nabla_{\nu} h^{\lambda}_{\,\lambda} - 
			\nabla_{\lambda} \nabla^{\lambda} h_{\mu\nu}\right) \\ \label{deltaR}
	\delta R & = & - h^{\mu\nu} \, R_{\mu\nu} + \nabla_{\mu} \left( \nabla_{\nu} h^{\mu\nu} - \nabla^{\mu} h^{\nu}_{\,\nu}\right)
\end{eqnarray}

In section \eqref{sec:GSHJeqn} we express the H-J functional for the gravity-scalar system in terms of 
quantities defined on constant $\tau$ hypersurfaces, like the intrinsic curvature $\RR$ and the extrinsic curvature $K_{\mu\nu}$. We need to vary these quantities in order to calculate the momenta conjugate to $\vf$ and $g_{ij}$. 
\begin{eqnarray}
	\delta n_{\mu} & = & \frac{1}{2} \, h_{\mu}^{\,\,\,\nu} n_{\nu}  \\
	\delta K_{\mu\nu} & = & \frac{1}{4} \left( h_{\nu}^{\,\,\,\lambda} K_{\mu\lambda} + h_{\mu}^{\,\,\,\lambda} K_{\nu\lambda} \right) +
		\frac{1}{4} \, n_{\lambda} \left( \nabla_{\nu} h^{\lambda}_{\,\mu} + \nabla_{\mu} h^{\lambda}_{\,\nu} - 
			2 \nabla^{\lambda} h_{\mu\nu}\right) \\
	\delta K & = & - \frac{1}{2} \, h^{\mu\nu} K_{\mu\nu} + \frac{1}{2} \, n_{\mu} \, \left( \nabla_{\nu} h^{\mu\nu} 
			- \nabla^{\mu} h^{\nu}_{\,\nu} \right)
\end{eqnarray}
The metric induced on the hypersurfaces, $\tg_{\mu\nu}=g_{\mu\nu}+n_{\mu} n_{\nu}$, is just the spatial part of the metric, $g_{ij}$.
Using the expression for $\delta n_{\mu}$ we can verify that $\delta \tg_{\mu\nu}$ is also purely spatial:
\begin{eqnarray}
	\delta \tg_{\mu\nu} & = & h_{\mu\nu} \, \delta_{\mu i} \, \delta_{\nu j}
\end{eqnarray}
The variation of the intrinsic curvature tensors can be read off from \eqref{deltaRmunu} and \eqref{deltaR} by replacing all four dimensional indices $\mu,\nu,\ldots$ with spatial indices $i,j,\ldots$, and replacing the covariant derivatives $\nabla_{\mu}$ with the covariant derivatives for the spatial metric, $D_i$.

\subsection{Varying Einstein's Equation}

In section \eqref{sec:slowroll} we need the first order variation of 
Einstein's equation:
\begin{eqnarray}
	H_{\mu\nu} & = & \delta G_{\mu\nu} - 8 \pi G \, \delta T_{\mu\nu}
\end{eqnarray}
Recall that:
\begin{eqnarray}
	G_{\mu\nu} & = & R_{\mu\nu} - \frac{1}{2} \, g_{\mu\nu} \,R \\
	T_{\mu\nu} & = & \nabla_{\mu} \vf \nabla_{\nu} \vf - g_{\mu\nu} \, \left( \frac{1}{2} \, \nabla^{\lambda} \vf \nabla_{\lambda} \vf + V(\vf)\right)
\end{eqnarray}
We can then use the results of the previous section to write out the first order variation of these expressions:
\begin{eqnarray}
	\delta G_{\mu\nu} & = & \frac{1}{2} \, \left( \nabla_{\lambda} \nabla_{\mu} h^{\lambda}_{\,\nu} + 
			\nabla_{\lambda} \nabla_{\nu} h^{\lambda}_{\,\mu} - \nabla_{\mu} \nabla_{\nu} h^{\lambda}_{\,\lambda} - 
			\nabla_{\lambda} \nabla^{\lambda} h_{\mu\nu}\right) - \frac{1}{2} \, h_{\mu\nu} \, R \\ \nonumber
			& & \hspace{10 pt} +\frac{1}{2} \, g_{\mu\nu} \, h^{\lambda \rho} R_{\lambda \rho} - \frac{1}{2} \, g_{\mu\nu} \nabla_{\lambda}
				 \left( \nabla_{\rho} h^{\lambda \rho} - \nabla^{\lambda} h^{\rho}_{\,\rho}\right) \\
	\delta T_{\mu\nu} & = & \nabla_{\mu} \vf \nabla_{\nu} \dvf + \nabla_{\mu} \dvf \nabla_{\nu} \vf -g_{\mu\nu} \left( \nabla^{\lambda} \vf
			\nabla_{\lambda} \dvf + \pvf V \, \dvf \right)  \\ \nonumber
			& & - \hspace{10pt} h_{\mu\nu} \left( \frac{1}{2} \, \nabla^{\lambda} \vf \nabla_{\lambda} \vf + V\right) + \frac{1}{2} \, g_{\mu\nu} \, 
				h^{\lambda \rho} \nabla_{\lambda} \vf \nabla_{\rho} \vf
\end{eqnarray}
In section \eqref{sec:slowroll} we keep only the scalar degrees of freedom and choose the
longitudinal gauge, {\it i.e.} we write the metric as \eqref{scalmet} with $E=B=0$.
Then the variations take the explicit form:
\begin{eqnarray}
	H_{\tau \tau} & = & -2 a^2 \vec{D}^{\,2} \, \psi + 6 \HH \psi\,' - 8 \pi G \, a^2 \, \chi \, \pvf V \\
	H_{ij} & = & - \delta_{ij} \left( 2 \,\psi\,'' + a^2 \vec{D}^{\,2} (\zeta - \psi) + 2 \,(\HH^2 + \HH\,')(\zeta+\psi)  + 2 \,\HH \,(\zeta\,' + 2 \psi \,' )\right) \\
	\nonumber & & \hspace{10pt} + 8 \pi G \, \delta_{ij} \left( a^2 \chi \, \pvf V - \vf\,' \chi \,' - (\vf\,')^2 \, (\zeta + \psi)\right) 
		- \partial_i \, \partial_j \, \left( \zeta - \psi \right) \\
	H_{\tau i} & = & - 2 \partial_i \, \left( \psi\,' + \HH \, \zeta + 4 \pi G \, \vf\,' \chi \right)
\end{eqnarray}
The equations of motion are 
\begin{eqnarray}
	H_{\mu\nu} & = & 0
\end{eqnarray}
for all these components. In longitudinal gauge two of these equations are constraints. 
$H_{\tau i} = 0$ relates $\zeta$ and $\chi$ as \eqref{zetachiconstraint} and 
$H_{ij}=0$ with  $i \neq j$: 
\begin{eqnarray}
	\partial_i \, \partial_j \, \left( \zeta - \psi \right) & = & 0 
\end{eqnarray}
clearly gives $\zeta=\psi$. 

\subsection{Variations of the Momenta}
\label{momenvar}
The H-J functional to quadratic order in the fluctuations is computed in section \eqref{sec:slowroll}
from the variation of different expressions for the canonical momenta. We need the
variation of the usual expressions for canonical momenta:
\begin{eqnarray}
	\pi_{\vf} & = & - n_{\mu} \nabla^{\mu} \vf \\
	\pi_{ij} & = & \frac{1}{2} \, \left( K_{ij} - g_{ij} K \right) 
\end{eqnarray}
They are:
\begin{eqnarray}
	\delta \pi_{\vf} & = &  - n_{\mu} \nabla^{\mu} \chi + \frac{1}{2} \, h^{\mu\nu} n_{\mu} \nabla_{\nu} \vf \\
          \delta \pi_{ij} & = & \frac{1}{8} \, \left( h_{j}^{\,\,k} K_{ik} + h_{i}^{\,\,k} K_{jk}\right) -
            \frac{1}{2} \, h_{ij} \, K + 
		\frac{1}{4} \, g_{ij} \, h^{\lambda \rho} K_{\lambda \rho} \\ \nonumber
				&   & + \frac{1}{8} \, n_{\lambda} \left( \nabla_{j} h^{\lambda}_{\,i} + \nabla_{i} h^{\lambda}_{\,j} - 2 \nabla^{\lambda} h_{ij}\right) 
					- \frac{1}{4} \, g_{ij} n_{\lambda} \left( \nabla_{\rho} h^{\lambda \rho} - \nabla^{\lambda} h^{\rho}_{\,\rho} \right)
\end{eqnarray}
We also need the variations of the momenta derived from the local form of the H-J functional:
\begin{eqnarray}
	P_{\vf} & = & \pvf U - \pvf M \, \vec{D} \vf \cdot \vec{D} \vf - 2 M \vec{D}^{\,2} \vf + \pvf \Phi \, \RR \\
		P_{ij} & = & \frac{1}{2} \, g_{ij} \left( U + M \vec{D} \vf \cdot \vec{D} \vf \right) - M D_i \vf D_j \vf - \Phi \GG_{ij} + 
		D_i D_j \Phi - g_{ij} \vec{D}^{\,2} \Phi 
\end{eqnarray}
They are:
\begin{eqnarray}
			\delta P_{\vf} & = & \left( \pvf^{\,2} U - \pvf^{\,2} M \vec{D}\vf \cdot \vec{D}\vf - 2 \pvf M \vec{D}^{\,2} \vf + \pvf^{\,2} \Phi \, \RR \right) \, \chi 
			- 2 \pvf M \vec{D} \vf \cdot \vec{D} \chi \\  \nonumber
			& & - 2 M \vec{D}^{\,2} \chi + \pvf M \, h^{ij} D_i \vf D_j \vf + 2 M h^{ij} D_i D_j \vf + 2 M D_i h^{ij} D_j \vf  \\ \nonumber
			& &- M \vec{D} \vf \cdot \vec{D} h^{j}_{\,j} - \pvf \Phi h^{ij} \RR_{ij} + \pvf \Phi \, D_i \left( D_j h^{ij} - D^i h^{j}_{\,j}\right) \\
	\delta P_{ij} & = & \frac{1}{2} \, h_{ij} \left( U + M \vec{D} \vf \cdot \vec{D} \vf \right) - \frac{1}{2} \, M \, h^{kl} D_k \vf D_l \vf
		- h_{ij} \vec{D}^{\,2} \Phi + g_{ij} \, h^{kl} D_k D_l \Phi \\ \nonumber
			& & - \frac{1}{2} \, \left( D_k D_i h^{k}_{\,j} + D_k D_j h^{k}_{\,i} - D_i D_j h^{k}_{\,k} - \vec{D}^{\,2} h_{ij} 
				- g_{ij} D_k D_l h^{kl} + g_{ij} \vec{D}^{\,2} h^{k}_{\,k} - h_{ij} \RR + g_{ij}\, h^{kl} \RR_{kl}\right) \\ \nonumber
			& & -\frac{1}{2} \, D_k \Phi \, \left( D_i h^{k}_{\,j} + D_j h^{k}_{\,i}- D^k h_{ij} \right) + 
			\frac{1}{2} \, g_{ij} D_k \Phi \, \left( D_l h^{kl} - D^k h^{l}_{\,l} \right) 
\end{eqnarray}
The results stated here are completely general. In \eqref{varPvf} to \eqref{varpij}
they are specialized to spatially flat FRW backgrounds.

%\pagebreak

\end{document}